\documentclass[journal]{vgtc}
\usepackage{fancyhdr}
\fancypagestyle{preprint}{%
  \fancyhf{}%
  \fancyhead[C]{\small\itshape Submitted to IEEE Transactions on Visualization and Computer Graphics, 18 May 2026}%
}

\usepackage[utf8]{inputenc}
\usepackage[T1]{fontenc}
\usepackage{graphicx}
\usepackage{amsmath,amssymb}
\usepackage{booktabs}
\usepackage{xcolor}
\usepackage{enumitem}
\usepackage{microtype}
\usepackage{balance}
\usepackage{makecell}
\usepackage{pifont}

\vgtccategory{Research}

\title{SmartIterator: Visual Analytics Workflows for Supervising Unsupervised Data Grouping}

\author{
  \authororcid{Gennady Andrienko}{0000-0002-8574-6295}%
  \thanks{This work was supported by Federal Ministry of Education and Research of Germany and the state of NRW as part of the \emph{Lamarr Institute for Machine Learning and Artificial Intelligence}}~~and  
  \authororcid{Natalia Andrienko}{0000-0003-3313-1560}
}
\authorfooter{
  \item Gennady and Natalia Andrienko are with Fraunhofer Institute IAIS, Sankt Augustin, Germany, the Lamarr Institute for Machine Learning and Artificial Intelligence, Sankt Augustin, Germany, and City St~George's, University of London, UK.
}

\teaser{
  \centering
  \begin{minipage}[c]{0.33\textwidth}
    \begin{subfigure}{\textwidth}
      \centering
      \includegraphics[width=\textwidth]{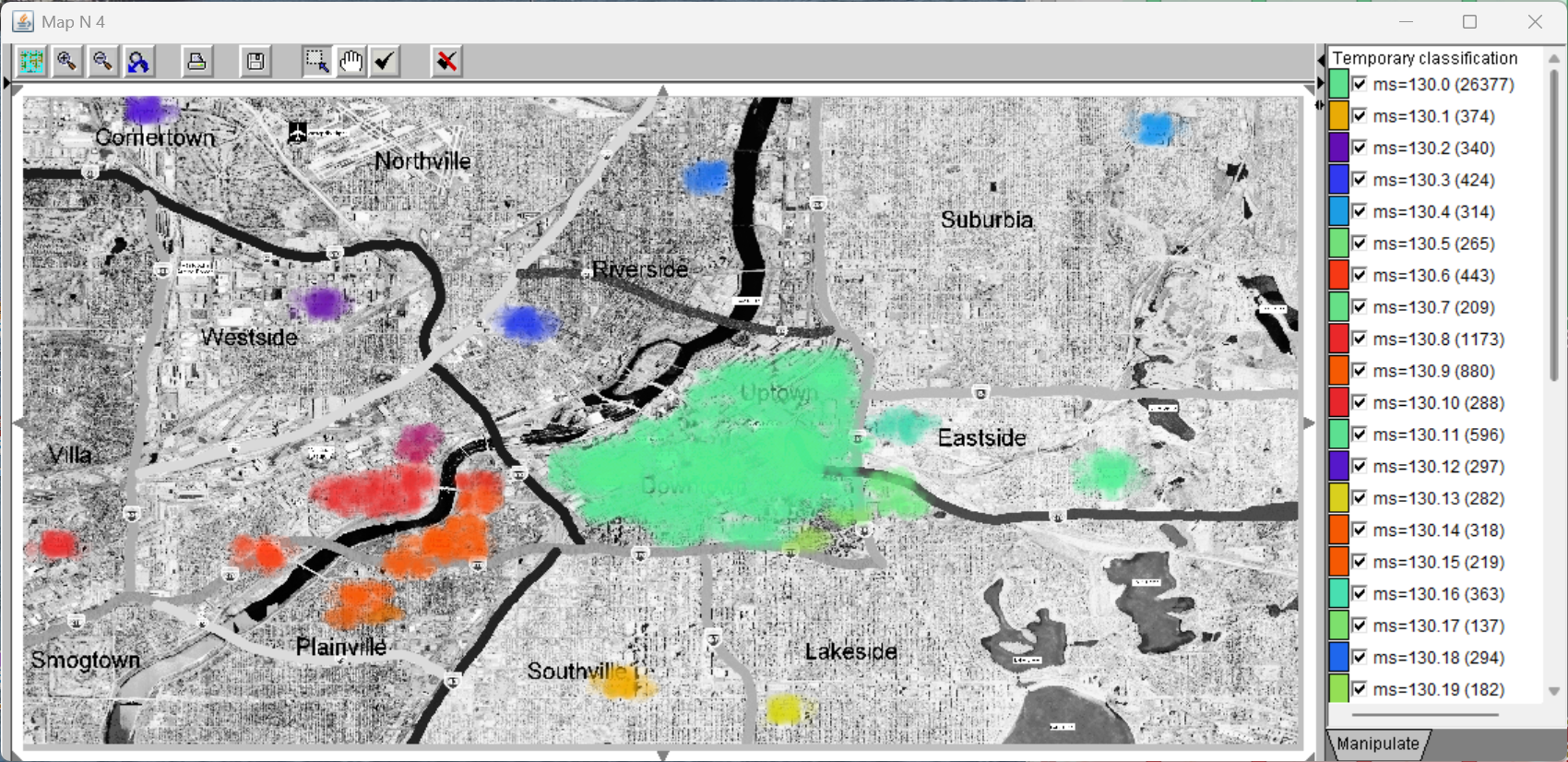}
      \caption{Spatial distribution of 26 clusters.}
      \label{fig:map}
    \end{subfigure}

    \vspace{1pt}

    \begin{subfigure}{\textwidth}
      \centering
      \includegraphics[width=\textwidth]{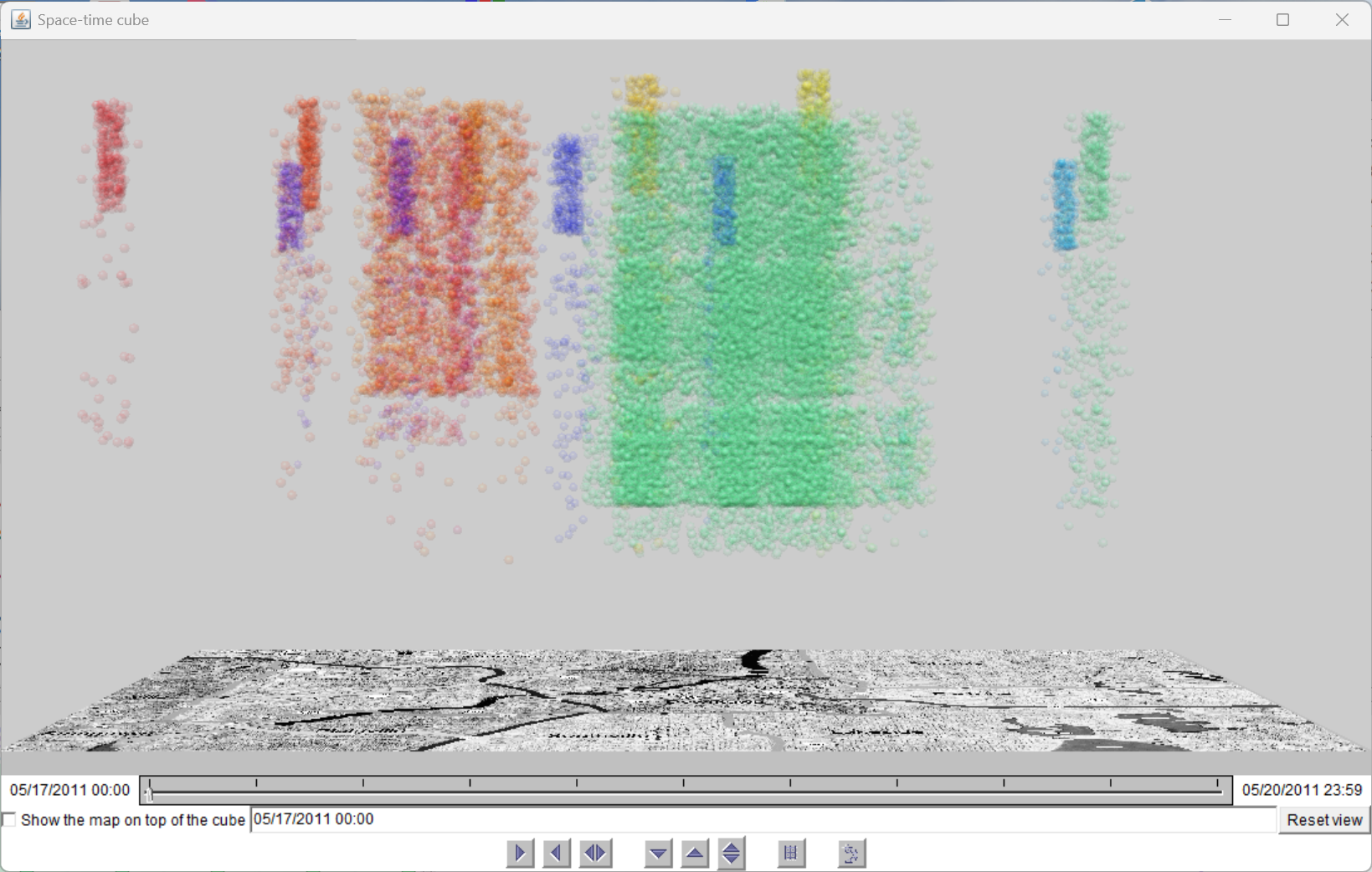}
      \caption{Space-time cube showing temporal dynamics.}
      \label{fig:stcube}
    \end{subfigure}
  \end{minipage}%
  \hfill
  \begin{minipage}[c]{0.66\textwidth}
    \begin{subfigure}{\textwidth}
      \centering
      \includegraphics[width=\textwidth]{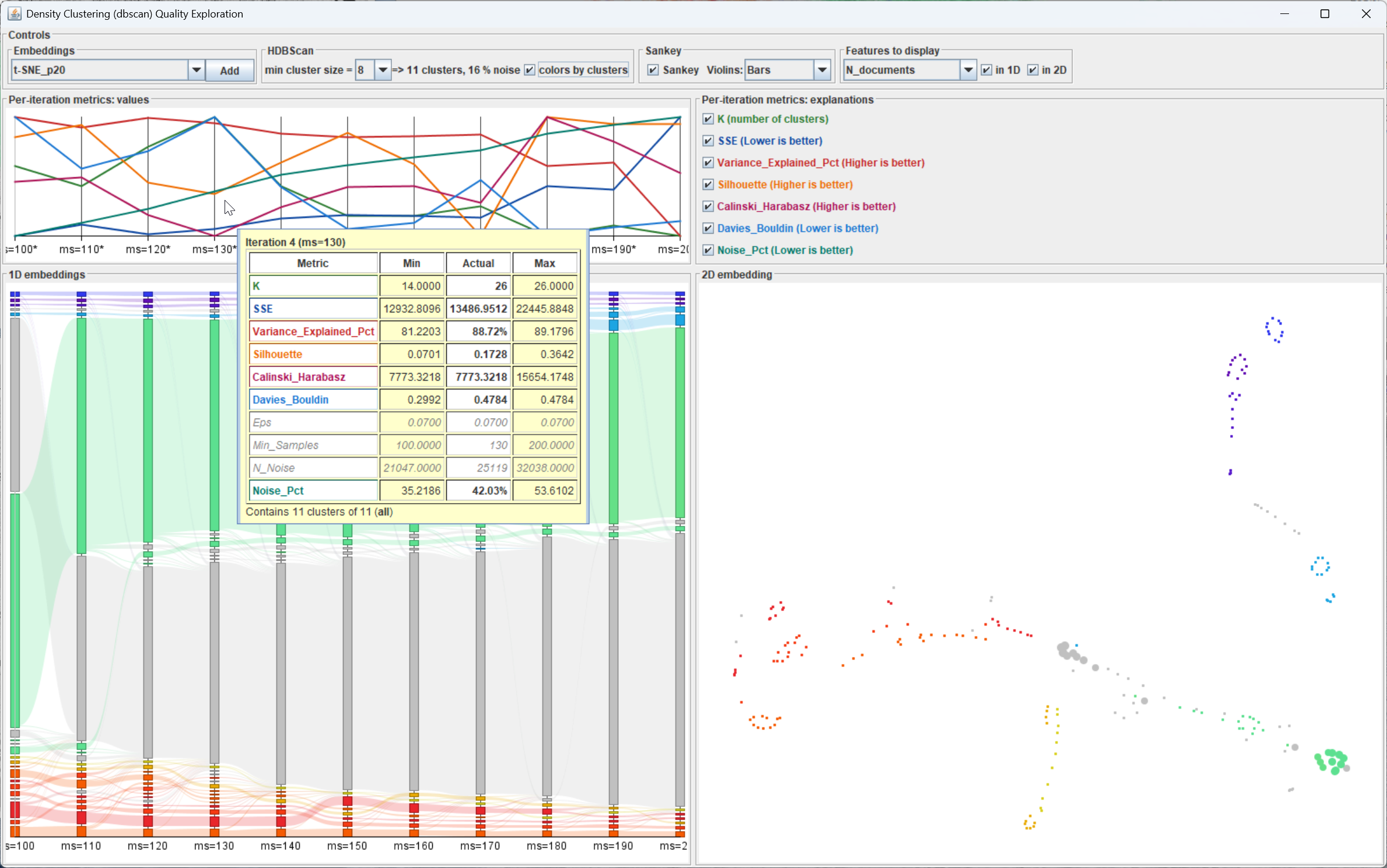}
      \caption{IteraScope display}
      \label{fig:dbscan}
    \end{subfigure}
  \end{minipage}

  \caption{SmartIterator applied to density-based clustering of social-media messages across a \texttt{min\_samples} sweep.
    The IteraScope display~(c) combines a 1D group embedding with Sankey transitions (center), a quality-metrics chart (top), and a 2D embedding with HDBSCAN-detected recurrent archetypes (right); iterations containing all archetypes are highlighted as candidates.
    Linked domain views---map~(a) and space-time cube~(b)---contextualize and validate groupings, completing the cycle of supervising the unsupervised.}
  \label{fig:overview}

  \label{fig:teaser}
}

\abstract{%
Unsupervised learning methods---topic modeling, partition-based and density-based clustering---produce data groupings without human guidance, yet choosing and evaluating those groupings should not itself be unsupervised.
We present \emph{SmartIterator}~(SI), a visual analytics approach that treats the full sequence of grouping results across a parameter sweep as a first-class analytical object.
For each method family, SI provides a structured six-phase workflow that guides the analyst through systematic exploration of grouping results---from quality-metric overview through transition-stability assessment, membership-confidence evaluation, content and context inspection, and recurrent-archetype verification to an informed decision---building cumulative understanding of data structure along the way.
The workflows are operationalized through \emph{IteraScope}~(IS), a coordinated visual display combining quality-metric charts with semantic color encoding, a 1D group embedding with Sankey-style transition flows and violin plots of membership confidence, a 2D group embedding with HDBSCAN-detected recurrent archetypes that highlights iterations capturing all persistent patterns, and domain-specific linked views for contextualized interpretation.
We demonstrate the three workflows on:
(1)~simulated social-media messages from the VAST Challenge 2011 (density-based clustering, validated against ground truth),
(2)~EU population statistics across ${\sim}1\,500$ NUTS-3 regions (partition-based clustering), and
(3)~30 years of IEEE VIS papers (NMF topic modeling).
The workflows constitute the main contribution: they provide actionable, method-specific guidance for navigating parameter spaces, studying how data structure evolves across configurations, and grounding analytical understanding in domain context---yielding knowledge about the data that no single ``best'' result can provide.
}


\begin{document}
\firstsection{Introduction}
\maketitle
\thispagestyle{preprint}

For many years we have observed students, researchers, and practitioners selecting the number of clusters or topics by examining a single summary statistic---a silhouette score, an SSE elbow, a coherence curve---and picking the ``best'' value.  This ritual reduces a rich, multi-faceted structural question to a one-dimensional optimization problem, discarding everything the remaining configurations could reveal about data structure, group stability, and analytical confidence.

Unsupervised learning methods are indispensable for discovering structure in data, yet their results are highly sensitive to parameter choices. The number of topics or clusters~$K$, the neighborhood radius~$\varepsilon$, the minimum number of neighbors, random seeds---each configuration can yield a substantially different grouping of the same data~\cite{vonluxburg2010clustering, bendavid2006sober}. We argue that the resulting question---\emph{which groups are ``real,'' and what does the data's structure look like?}---is best answered not by seeking a single optimum but by \emph{supervising} the unsupervised process: systematically exploring the sequence of grouping results, studying how structure emerges and transforms across configurations, and building domain-grounded understanding that transcends any individual solution.

Existing approaches address fragments of this challenge. Stability metrics~\cite{vonluxburg2010clustering, hennig2007cluster} compress complex structural information into single numbers. Ensemble methods~\cite{strehl2002cluster, fred2005combining, monti2003consensus} aggregate multiple runs into a consensus partition, discarding individual solutions and inter-solution relationships. Visual analytics tools such as ClusterVision~\cite{kwon2018clustervision}, Clustrophile~2~\cite{cavallo2019clustrophile}, and Termite~\cite{chuang2012termite} support exploration or pairwise comparison of individual results but do not track how groups evolve across a \emph{sequence} of configurations. Crucially, none offers \textbf{structured, method-specific workflows} that guide an analyst from an initial parameter sweep to a justified final choice.

We present \emph{SmartIterator}~(SI), a visual analytics approach for supervising unsupervised data grouping, operationalized through \emph{IteraScope}~(IS), a coordinated visual display. Our contributions are:

\begin{itemize}
\item \textbf{Three parameter-supervision workflows}---for density-based clustering, partition-based clustering, and topic modeling---each a structured six-phase sequence that guides the analyst through systematic exploration of grouping results, building cumulative understanding of data structure through stability assessment, confidence evaluation, archetype verification, and domain-contextualized interpretation;
\item \textbf{IteraScope}, a coordinated visual display combining quality-metric charts with semantic color encoding, a 1D group embedding with Sankey-style transition flows and violin plots of membership confidence, a 2D group embedding with HDBSCAN-based detection of recurrent group archetypes and interactive threshold control, word clouds and prominent-term tooltips, and domain-specific linked views;
\item \textbf{Demonstration on three datasets} representing different data types and method families---VAST Challenge 2011 synthetic social-media messages (density-based clustering, validated against ground truth), EU population statistics across ${\sim}1\,500$ NUTS-3 regions (partition-based clustering), and 30 years of IEEE VIS papers (NMF topic modeling).
\end{itemize}

The remainder of this paper is organized as follows. Section~\ref{sec:related} reviews related work. Section~\ref{sec:tasks} presents analytical goals and tasks. Section~\ref{sec:overview} gives an overview of the SI approach and the IS display. Section~\ref{sec:computation} details the computational components. Section~\ref{sec:visual} introduces IS components progressively through three workflow demonstrations---one per method family---followed by a consolidated reference of all components and interactions. Section~\ref{sec:workflows} distills the three SI workflows as the core contribution. Sections~\ref{sec:evaluation} and~\ref{sec:discussion} present evaluation and discuss limitations. Section~\ref{sec:conclusion} concludes. 

Supplementary HTML Appendices~A--C provide full-resolution figures and additional analysis steps for each demonstration; three Google Colab notebooks supply a reproducible Python implementation of the workflow.\footnote{All supplementary materials are available at \url{https://geoanalytics.net/and/SmartIterator}.}

\section{Related Work}
\label{sec:related}

Our work draws on four areas: visual analytics for unsupervised learning, visual comparison and tracking of groupings, clustering stability theory and ensemble methods, and domain-specific visual analytics combined with dimensionality reduction. We review each in turn, positioning SmartIterator relative to the closest prior art.

\subsection{Visual Analytics for Unsupervised Learning}
\label{sec:rw_va_unsupervised}

Sedlmair et al.~\cite{sedlmair2014visual} provide a conceptual framework for visual parameter space analysis, identifying strategies---sensitivity analysis, local navigation, global overview---that analysts use to navigate high-dimensional parameter spaces. SI instantiates several of these strategies for unsupervised-learning parameters, adding method-specific workflows that link parameter choices to group-level stability evidence.

ClusterVision~\cite{kwon2018clustervision} ranks clustering results by quality metrics and supports side-by-side comparison of individual solutions. Clustrophile~2~\cite{cavallo2019clustrophile} offers a guided workflow for iterative clustering refinement, including projection-based exploration and feature selection. Both tools help analysts select among individual solutions but do not track how \emph{specific groups} evolve across the parameter space---a key requirement for stability assessment. 

In the topic-modeling domain, Termite~\cite{chuang2012termite} provides a matrix view of term--topic distributions, TopicLens~\cite{kim2017topiclens} supports efficient multi-level exploration of large document collections, and LDAvis~\cite{sievert2014ldavis} supports interactive exploration of a single model through inter-topic projection. 
El-Assady et al.~\cite{elassady2017progressive} propose progressive topic-model refinement with visual feedback but focus on a single evolving model rather than comparison across configurations. 
Chuang et al.~\cite{chuang2012interpretation} study interpretation and trust in model-driven text visualizations, establishing design principles for making topic-model outputs transparent. Choo et al.~\cite{choo2013utopian} introduce \textsc{Utopian}, an interactive NMF-based topic-modeling system that enables human-in-the-loop refinement of topics---the closest prior art in terms of computational method (also NMF, also interactive), though limited to a single model without cross-configuration comparison or stability tracking.
Wenskovitch et al.~\cite{wenskovitch2018towards} combine dimensionality reduction and clustering in a human-in-the-loop setting, enabling analysts to steer a single result through projection interactions---but without cross-configuration comparison or stability assessment.

Pister et al.~\cite{pister2021pkclustering} introduce PK-clustering, a mixed-initiative approach in which the analyst specifies prior knowledge as partial cluster seeds, an ensemble of algorithms is executed and ranked by edit-distance match to the seeds, and a final partition is consolidated through consensus inspection. PK-clustering and SI share the premise that multiple unsupervised results should be juxtaposed for analyst-driven reconciliation rather than collapsed into a single consensus. However, PK-clustering varies the \emph{algorithm} at fixed parameters on a single graph-clustering task and treats the result set as unordered, whereas SI varies a \emph{parameter} within a chosen method family and treats the sequence as a first-class analytical object whose ordering encodes structural evolution; it provides no transition-flow tracking, membership-confidence assessment, or recurrent-archetype detection across configurations.

More broadly, Stolper et al.~\cite{stolper2014progressive} formalize \emph{progressive visual analytics}, advocating that systems expose intermediate results for user-driven exploration---a principle SI realizes by exposing every iteration for interactive inspection within a structured, six-phase analytical workflow.
Cashman et al.~\cite{cashman2019user} propose a user-based workflow for exploratory model analysis that externalizes model-comparison decisions; SI specializes this concept for the iterative parameter-selection problem by providing structured, method-specific workflows with group-level stability evidence and transition tracking.

Unlike these tools, SI treats the full \emph{sequence} of grouping results as a first-class analytical object and provides method-specific workflows---operationalized through the IS display---that guide the analyst through systematic exploration building domain-grounded understanding of data structure, culminating in a justified, well-informed decision.

\subsection{Comparing and Tracking Groupings}
\label{sec:rw_comparison}


Gleicher et al.~\cite{gleicher2011visual} identify three strategies for visual comparison---juxtaposition, superposition, and explicit encoding---and argue that effective designs combine them. SI leverages all three in concert: iterations are juxtaposed as parallel 1D axes that preserve parameter ordering, groups from different iterations are superposed in the shared 2D embedding to reveal spatial correspondence, and Sankey-style bands explicitly encode membership transitions between consecutive configurations. This combination lets analysts see both the overall trajectory of structural change and the fine-grained document flows that drive it. Alsallakh et al.~\cite{alsallakh2012reinventing} take a complementary approach, visualizing cluster-to-cluster membership overlaps through contingency wheels that foreground set-theoretic relationships; our Sankey encoding trades the symmetric contingency view for a directed, flow-volume representation that emphasizes sequential evolution across the parameter axis.

Yu et al.~\cite{yu2026parallel} introduce ParaClus, a parallel-axes system for comparing cluster structures derived from different embeddings and attributes. Each axis represents a partition, and inter-axis connections reveal data-item overlaps---a layout closely related to our 1D group embedding with Sankey transitions. However, ParaClus targets cross-model comparison of static embeddings rather than tracking how groups evolve along a directed parameter sequence; it does not treat the parameter axis as an ordered sequence whose transitions encode structural evolution, and provides no archetype-based stability signal or membership-confidence assessment.

Alexander and Gleicher~\cite{alexander2016task} develop a task-driven approach for comparing topic models at different~$K$, providing the closest precedent for the cross-$K$ comparison task in our topic-modeling workflow. We extend their work by supporting additional method families, adding transition-flow visualization that tracks splits and merges, incorporating membership-confidence assessment via violin plots, and linking to domain-specific views for contextualized interpretation.

Alluvial diagrams~\cite{rosvall2010mapping} for tracking community evolution in networks directly inspire our Sankey transitions between iterations. Interactive Sankey diagrams~\cite{riehmann2005interactive} and Parallel Sets~\cite{kosara2006parallel} provide foundational interaction techniques that we adapt to parameter exploration. TextFlow~\cite{cui2011textflow} uses river-like flows to track topic evolution over \emph{time}; our flows track group evolution over \emph{parameter configurations}---a fundamentally different analytical axis where ordering reflects increasing model complexity rather than temporal progression.

\subsection{Clustering Stability and Ensembles}
\label{sec:rw_stability}

Von~Luxburg~\cite{vonluxburg2010clustering} formalizes the intuition that a ``good'' number of clusters yields results robust to perturbations. Campello et al.~\cite{campello2013density} propose hierarchical density-based clustering (HDBSCAN), which extracts clusters at varying density thresholds from a single hierarchy; McInnes et al.~\cite{mcinnes2017hdbscan} provide the scalable open-source implementation that SI employs for both the density-based clustering workflow and the archetype-detection mechanism. Ben-David et al.~\cite{bendavid2006sober} provide a formal definition of stability and discuss its relationship to the true number of clusters. Hennig~\cite{hennig2007cluster} extends the framework to cluster-wise stability via bootstrap resampling, enabling per-group stability scores---a concept that our Sankey transitions and archetype detection operationalize visually.

Ensemble methods aggregate multiple results into a single consensus: evidence accumulation~\cite{fred2005combining} builds a co-association matrix; cluster ensembles~\cite{strehl2002cluster} define consensus functions optimizing agreement across partitions; consensus clustering~\cite{monti2003consensus} uses resampling-based approaches. While these methods produce a single ``best'' partition, they discard individual solutions and the transitions between them. 

SI preserves individual results and their inter-relationships for interactive exploration. Where ensemble methods answer ``what is the consensus?'', SI answers ``how does structure evolve across configurations, and where is it stable?''---and what domain knowledge can be extracted from studying that evolution.
Recent neural topic models such as BERTopic~\cite{grootendorst2022bertopic} introduce additional parametric choices (embedding model, UMAP parameters, HDBSCAN thresholds); extending stability assessment to such pipelines is a natural direction that our architecture already partially supports through its method-agnostic iteration structure.

\subsection{Domain-Specific Analytics and Dimensionality Reduction}
\label{sec:rw_domain_dr}

Andrienko and Andrienko~\cite{andrienko2006exploratory} provide a systematic framework for exploratory analysis of spatial and temporal data whose task taxonomies inform our domain-linked views. Ferreira et al.~\cite{ferreira2013visual} combine clustering with map-based exploration for urban data, showing that geographic coherence is a powerful external validator of cluster quality---a principle we exploit in our partition-clustering workflow. Guo et al.~\cite{guo2006} couple self-organizing maps with linked geographic displays, establishing a computation--visualization integration pattern that SI extends to the multi-configuration setting. Our framework adds iterative stability assessment to these domain-specific workflows, bridging parameter-space exploration with domain-contextualized interpretation.

Nonato and Aupetit~\cite{nonato2019multidimensional} survey multidimensional projection techniques, linking them with distortions and tasks. Sedlmair et al.~\cite{sedlmair2012taxonomy} provide a taxonomy of visual cluster separation factors that informs how well projected layouts communicate group structure. Wang et al.~\cite{wang2021understanding} empirically compare DR methods including PaCMAP, establishing the trade-offs between local and global structure preservation that guide our projection choices for group-level embeddings. Espadoto et al.~\cite{espadoto2021toward} provide a quantitative benchmark informing the choice among alternatives. We apply 1D projections (user-selectable: t-SNE, UMAP, PaCMAP, or LocalMAP) to order groups along Sankey axes and 2D projections for scatter-plot embedding---in both cases on \emph{group-level} feature vectors, so that proximity encodes cross-iteration group similarity. 

Sacha et al.~\cite{sacha2014knowledge} frame visual analytics as a loop of hypothesis generation, evidence gathering, and insight formation; SI implements this loop through automated computation feeding interactive exploration, which in turn informs parameter refinement. The broader agenda of visual analytics for human-centered machine learning~\cite{andrienko2022humancentered, keim2008visual} motivates our focus on making unsupervised outputs transparent, comparable, and interpretable through structured workflows.

\section{Analytical Goals and Tasks}
\label{sec:tasks}

Our task framework derives from years of applying unsupervised learning to spatio-temporal, textual, and multivariate tabular data. The central challenge is not merely computational (finding a grouping) but \emph{analytical}: understanding what the data's structure looks like, how it changes across parameter configurations, and what domain knowledge can be extracted from the full landscape of grouping results. We structure this challenge into three goals and eight tasks.

\subsection{Goals}

\begin{description}[nosep,leftmargin=*,font=\normalfont\bfseries]
  \item[G1\,---\,Assess grouping stability.] Determine which groups persist across parameter configurations and which are artifacts of a particular setting. Stability is a necessary (though not sufficient) condition for a group to reflect genuine data structure.
  \item[G2\,---\,Understand group composition and meaning.] Examine the members, feature distributions, representative content, and domain context of individual groups to judge their interpretive value.
  \item[G3\,---\,Build structural understanding.] Develop a comprehensive understanding of how data structure evolves across configurations---which patterns persist, where transitions occur, and what domain meaning emerges---culminating in an informed parameter choice grounded in converging evidence from quality metrics, transition stability, membership confidence, archetype completeness, and domain context. The goal is not a single ``optimal'' answer but deep structural understanding that supports defensible analytical conclusions.
\end{description}

\subsection{Tasks}

The three goals decompose into eight analytical tasks, each linked to one or more goals. Together they define the analytical space that the SI workflows (Section~\ref{sec:workflows}) navigate and the IS display (Section~\ref{sec:visual}) supports.

\begin{description}[nosep,leftmargin=*,font=\normalfont\bfseries]
  \item[T1\,---\,Compare quality across iterations.] Assess how quality metrics (silhouette, coherence, Davies--Bouldin, noise fraction, etc.) evolve across the parameter range. Identify elbows, peaks, plateaus, and anomalies that narrow the candidate range. \emph{(G3)}
  \item[T2\,---\,Track group transitions.] Observe how groups split, merge, grow, and shrink between configurations. Splits may indicate meaningful sub-structure; absorptions may indicate over-merging. \emph{(G1)}
  \item[T3\,---\,Identify persistent groups.] Find groups that remain cohesive---neither splitting nor merging substantially---across many iterations. Wide-range persistence is strong evidence of genuine structure. \emph{(G1)}
  \item[T4\,---\,Examine group members and features.] Inspect composition: feature distributions for clusters, top-weighted terms and word clouds for topics, distance-to-centroid profiles, and representative items. \emph{(G2)}
  \item[T5\,---\,Contextualize in domain dimensions.] Relate groups to spatial locations (maps), temporal patterns (space-time cubes), textual content (word clouds), or multivariate profiles (parallel coordinates). Domain coherence is a powerful external validator of group quality. \emph{(G2)}
  \item[T6\,---\,Propagate groupings to linked views.] Export a selected grouping---or a transition between groupings---as a class attribute so that all linked displays use group colors for rendering, filtering, and aggregation. \emph{(G1, G2, G3)}
  \item[T7\,---\,Assess within-group confidence.] Evaluate membership certainty and outlier prevalence via violin plots of per-item membership probabilities and outlier scores. A group with uniformly high membership is well-defined; dispersed or bimodal scores suggest conflated sub-populations. \emph{(G1, G2)}
  \item[T8\,---\,Identify complete iterations via recurrent archetypes.] Apply HDBSCAN to pooled group-level feature vectors to detect recurrent archetypes---patterns appearing repeatedly across configurations. Mark iterations containing all archetypes as ``complete'' candidates capturing the full diversity of persistent structure. \emph{(G1, G3)}
\end{description}

\section{Approach Overview}
\label{sec:overview}

The SI approach comprises two components: an \emph{iterative computation engine} (Python) and the \emph{IS display} (Java), embedded within V-Analytics~\cite{andrienko2013visual}, which provides domain-specific linked views (maps, space-time cubes, parallel coordinates).

The computation engine accepts input data and a parameter-range specification: method (topic modeling, partition- or density-based clustering), sweep parameter (e.g.\ $K = 5 \ldots 30$, or $\texttt{min\_samples} = 3 \ldots 50$), and optional secondary parameters (random seeds, stop-word lists, bigram count). Iterations are distributed across CPU cores via \texttt{ProcessPoolExecutor}; shared read-only structures (TF-IDF matrix, pre-computed distances) are built once and passed through the pool initializer. Each iteration produces four outputs: (i)~per-item assignments with uncertainty indicators (membership probability and outlier score), (ii)~per-group metrics (prevalence, distance statistics, and---for topic modeling---coherence, exclusivity, diversity), (iii)~per-iteration quality metrics (Table~\ref{tab:metrics}), and (iv)~group-level feature vectors (centroids, medoids, or topic--term distributions).

After all iterations complete, the engine pools group-level feature vectors into a single matrix, applies HDBSCAN~\cite{mcinnes2017hdbscan} to identify recurrent group archetypes (T8), and computes 1D and 2D embeddings for the IS display. Details are given in Section~\ref{sec:comp_hdbscan}.

The IteraScope (IS, Fig.~\ref{fig:teaser}c) presents computation outputs in a coordinated multi-view layout: a quality-metric chart, a 1D group embedding with Sankey-style transition flows and violin plots of membership confidence, a 2D group embedding, and word clouds with term-level tooltips. Selected groupings are propagated as class attributes to domain-specific linked views, enabling class- and change-based coloring, filtering, and aggregation.

The analyst follows a method-specific SI \textit{workflow} (Section~\ref{sec:workflows}) through six phases: \textbf{(1)~metric overview} including archetype completeness $\to$ \textbf{(2)~transition assessment} via Sankey flows $\to$ \textbf{(3)~confidence evaluation} via violin plots $\to$ \textbf{(4)~content and context inspection} via terms, word clouds, and domain views $\to$ \textbf{(5)~archetype verification} in the 2D embedding $\to$ \textbf{(6)~decision} or refinement. Each phase contributes to cumulative understanding of data structure; the analyst may revisit earlier phases as insight deepens, and the knowledge gained from studying transitions and configurations is often as valuable as the final parameter choice itself.

\section{Computational Components}
\label{sec:computation}

This section details the computation engine that produces the artifacts consumed by IS: the three supported method families, uncertainty measures, quality metrics, archetype detection, and transition/embedding computations.

\subsection{Iterative Unsupervised Learning}
\label{sec:comp_methods}

All three method families share a common execution pattern: the user specifies a parameter range; the engine creates one task per configuration; tasks execute in parallel; and each task writes a standardized set of output files.

For topic modeling, we use Non-negative Matrix Factorization (NMF)~\cite{lee1999nmf} to decompose a TF-IDF document--term matrix~$V$ into non-negative factors $W$ (documents\,$\times$\,topics) and $H$ (topics\,$\times$\,terms). The text corpus undergoes configurable preprocessing: stop-word removal, minimum document-frequency filtering, and injection of frequent bigrams as compound tokens (e.g.\ ``machine\_learning''). The TF-IDF matrix and a Gensim dictionary for coherence computation~\cite{roder2015coherence} are built once and shared across workers. The primary sweep parameter is~$K$; optionally, multiple random seeds can be swept for a fixed~$K$ to assess seed sensitivity. Each task outputs the $W$ and $H$ matrices, per-document topic assignments (argmax of~$W$), per-topic top-$N$ term lists with weights, and all quality and uncertainty indicators.

For Partition-based clustering, we use K-means (scikit-learn, Lloyd's algorithm) with configurable $K$ and seed ranges. The primary sweep is over~$K$ with a fixed seed; alternatively, seeds are swept for a fixed~$K$ to assess algorithmic stability. Each task outputs cluster labels, centroids, per-item distances to assigned centroids, and all quality and uncertainty indicators. Input features are standardized once before the sweep.

For Density-based clustering, the engine supports DBSCAN~\cite{ester1996dbscan}, OPTICS~\cite{ankerst1999optics}, and HDBSCAN~\cite{mcinnes2017hdbscan}. The primary sweep parameter depends on the method: $\varepsilon$ for DBSCAN, \texttt{max\_eps} for OPTICS, or \texttt{min\_cluster\_size} for HDBSCAN, with \texttt{min\_samples} held fixed---or vice versa. Unlike partition-based methods, density-based methods discover~$K$ from the data and may assign items to noise; the output therefore includes a noise group placed last. Medoids replace centroids as group representatives, since density-based clusters may have non-convex shapes.

Each parameter configuration constitutes an independent task, making the sweep embarrassingly parallel. Tasks are dispatched to a \texttt{ProcessPoolExecutor} for Parallel execution, whose pool size defaults to $\max(1,\;\text{CPU cores} - 1)$. Shared read-only structures---data matrix, TF-IDF matrix, Gensim dictionary, pre-computed distance matrices---are passed once via the pool initializer, avoiding redundant serialization.

\subsection{Uncertainty Measures}
\label{sec:comp_uncertainty}

A grouping assigns each item to exactly one group, but this assignment may be confident or borderline. To surface this distinction, every method produces two per-item indicators normalized to~$[0,1]$:

\begin{description}[nosep,leftmargin=*]
  \item[Membership probability] quantifies how clearly an item belongs to its assigned group. For \emph{K-means, DBSCAN, and OPTICS}, we use $(\text{silhouette}_i + 1) / 2$~\cite{rousseeuw1987silhouettes}; a value near~1 indicates unambiguous membership, near~0.5 a borderline item. For \emph{HDBSCAN}, we use the built-in persistence probability. For \emph{NMF topic modeling}, we use $\max(w_i) / \sum w_i$---the proportion of the document's topic mixture attributable to the dominant topic.
  \item[Outlier score] quantifies how atypical an item is within its assigned group. For \emph{K-means, DBSCAN, and OPTICS}, we use $d_i / \max_{j \in g} d_j$---distance to the centroid or medoid relative to the most distant member; a value near~0 indicates a core member, near~1 a peripheral item. For \emph{HDBSCAN}, we use the GLOSH outlier score. For \emph{NMF}, we use the normalized entropy $H(w_i) / \log K$; high entropy signals diffuse topic membership.
\end{description}

These two indicators are complementary: an item may have moderate membership probability yet a low outlier score, or vice versa. Their within-group distributions are visualized as violin plots in IS (Section~\ref{sec:vis_violin}).

\subsection{Quality Metrics}
\label{sec:comp_metrics}

SI computes quality metrics at two levels. \emph{Per-iteration metrics} (Table~\ref{tab:metrics}) summarize overall grouping quality; they feed the IS metrics chart and Phase~1 of the SI workflows. \emph{Per-group metrics} characterize individual groups: prevalence, distance statistics, and---for topic modeling---per-topic coherence ($C_V$)~\cite{roder2015coherence}, exclusivity, and diversity; these appear in tooltips and inform Phases~3--4.

\begin{table}[tb]
  \centering
  \caption{Per-iteration quality metrics by method family. Arrows indicate whether higher~($\uparrow$) or lower~($\downarrow$) values are preferred, determining the warm/cool color encoding in IS.}
  \label{tab:metrics}
  \small
  \begin{tabular}{lp{6.5cm}}
    \toprule
    \textbf{Family} & \textbf{Metrics} \\
    \midrule
    Topic modeling &
      Reconstruction~\%~$\uparrow$,\;
      Frobenius norm~$\downarrow$,\;
      global diversity~$\uparrow$,\;
      mean exclusivity~$\uparrow$,\;
      mean coherence~$C_V$~$\uparrow$,\;
      document sparsity~$\uparrow$,\;
      topic sparsity~$\uparrow$,\;
      silhouette~$\uparrow$
    \\[4pt]
    Partition-based &
      SSE~$\downarrow$,\;
      variance explained~\%~$\uparrow$,\;
      silhouette~$\uparrow$,\;
      Calinski--Harabasz~$\uparrow$,\;
      Davies--Bouldin~$\downarrow$
    \\[4pt]
    Density-based &
      SSE~$\downarrow$,\;
      variance explained~\%~$\uparrow$,\;
      silhouette~$\uparrow$,\;
      Calinski--Harabasz~$\uparrow$,\;
      Davies--Bouldin~$\downarrow$,\;
      noise~\%~$\downarrow$,\;
      $K$~discovered~(\emph{info})
    \\
    \bottomrule
  \end{tabular}
\end{table}

The metrics differ by family. For topic modeling, reconstruction~\% plays the role of SSE: both measure how well the model captures the data, with diminishing returns as~$K$ increases. Global diversity and mean exclusivity together detect redundant topics---a phenomenon specific to topic modeling with no direct clustering analogue. For density-based methods, noise~\% and discovered~$K$ track how structural decisions change with the sweep parameter.

\subsection{Recurrent Archetype Detection via HDBSCAN}
\label{sec:comp_hdbscan}

Quality metrics and transition flows assess \emph{local} stability (does a group persist to the next iteration?). To provide a \emph{global} stability signal, SI pools group-level feature vectors from all iterations into a single matrix and clusters them with HDBSCAN~\cite{mcinnes2017hdbscan}. This meta-clustering operates on group representatives---centroids, medoids, or $H$-row vectors---rather than individual data items.

Before processing, the pooled matrix is standardized column-wise, optionally reduced via PCA (retaining ${\geq}95\%$ variance), and L2-normalized to prevent scale differences or high-dimensional noise from dominating distances. Each HDBSCAN cluster represents a \emph{recurrent group archetype}---a pattern appearing in multiple iterations under different parameter settings. Groups assigned to HDBSCAN noise are idiosyncratic and unlikely to represent genuine structure. The number of archetypes depends on the \texttt{min\_cluster\_size} threshold, controllable via a spinner in IS; the default is $\lfloor N_{\text{iterations}} / 2 \rfloor$, reflecting the expectation that a genuine archetype appears in at least half the explored parameter configurations. Changing the threshold updates archetypes and complete iterations immediately.

An iteration containing representatives of \emph{all} detected archetypes is marked ``complete'': it captures the full diversity of recurrent structure. Complete iterations are highlighted in the metrics chart with dark gridlines and asterisk labels~(T8), directing attention to the most promising configurations.

Archetype membership drives color assignment. All group centers and archetype centroids are embedded into a shared 2D space using distance-preserving MDS; colors derive from 2D position via four-corner interpolation. Two modes are available: \emph{color by archetype} paints all groups of one archetype in its centroid color, making cross-iteration persistence visible; \emph{color by item} paints each group by its own position, so that atypical members of an archetype are visually flagged by color deviation.

Optionally, IS maps a user-selected group-level attribute to dot size in the 1D and 2D embeddings. Available attributes include HDBSCAN membership probability, group size (number of assigned items), and any computed per-group quality metric. The mapping is controlled via the IS interface. For example, sizing by membership probability causes groups closest to their archetype's core to appear large while peripheral or noise-assigned groups shrink; sizing by group size instead emphasizes dominant groups at each iteration. Combined with the dual color scheme, this encoding lets the analyst spot outlier or marginal groups at a glance without inspecting individual tooltips.

\subsection{Transition Computation and Group Embedding}
\label{sec:comp_transitions}

Between each pair of consecutive \emph{visible} iterations, we compute a document-overlap matrix: entry~$(i,j)$ counts items assigned to group~$g_i$ in iteration~$a$ and to group~$g_j$ in iteration~$b$. These counts determine Sankey band widths and feed transition-related tooltips.


Group-level feature vectors from all iterations form a single matrix---the same used for archetype detection. A user-selectable projection method (t-SNE~\cite{vandermaaten2008tsne}, UMAP~\cite{mcinnes2018umap}, PaCMAP~\cite{PaCMap2024}, or LocalMAP~\cite{LocalMAP2025}) is applied twice to this matrix---once targeting 1D to order groups on vertical axes so that similar groups across iterations appear at similar positions (making persistent structures apparent as horizontal Sankey bands) and once targeting 2D to place groups in a scatter plot for spatial reasoning about inter-group similarity. Color assignment is computed independently via MDS on the same matrix, ensuring that group colors remain stable regardless of which projection is selected for the display. Because MDS preserves global distance structure, groups that are semantically similar receive similar colors across all iterations, providing a consistent visual reference even when the analyst switches between projection methods for the 1D/2D layout.

\section{Visual Design and Workflow Demonstrations}
\label{sec:visual}

The IteraScope (IS) display is a coordinated multi-view panel that operationalizes the six-phase SI workflow. Rather than describing its components in isolation, we introduce them progressively through three demonstrations---one per method family---each exercising different IS components and interaction patterns. Section~\ref{sec:vis_density} demonstrates density-based clustering on VAST Challenge 2011 data, introducing the metrics chart, 1D/2D embeddings, Sankey transitions, and domain-linked views (map, space-time cube). Section~\ref{sec:vis_partition} demonstrates partition-based clustering on EU population data, adding parallel coordinates and transition-class propagation. Section~\ref{sec:vis_topics} demonstrates NMF topic modeling on IEEE VIS papers, adding word clouds, term-level tooltips, and temporal prevalence analysis. Section~\ref{sec:vis_summary} then provides a consolidated reference of all IS components and interactions. Supplementary HTML Appendices~A--C provide full-resolution figures and additional analysis steps for each demonstration; three Google Colab notebooks supply a reproducible Python implementation of the computational pipeline.

\subsection{Density-Based Clustering: VAST Challenge 2011}
\label{sec:vis_density}

The VAST Challenge 2011~\cite{vast2011} provides approximately 1\,000\,000 simulated microblog messages posted over roughly one month in the fictional city of Vastopolis. On one particular day, accidents released contaminants that spread through the city center via wind and along the river via water flow, causing widespread health complaints. Citizens reported symptoms---nausea, breathing difficulties, skin irritation---in their posts. On a map, spatial concentrations of sickness-related posts are visible near hospitals and in affected areas. These concentrations constitute a \emph{ground truth} that we aim to recover through density-based clustering of message locations, \emph{without} using message text or prior knowledge of the event.

This scenario is well suited for demonstrating human supervision. The sickness-related concentrations vary substantially in density: hospital clusters are tight, while wind- and water-dispersed complaints form elongated, diffuse patterns. Moreover, these concentrations are embedded in a background of unrelated messages that acts as spatial noise. No single parameter setting can simultaneously capture all concentrations at their native densities while rejecting the background: a tight $\varepsilon$ or high \texttt{min\_samples} recovers only the densest hospital clusters, while a loose setting absorbs noise. The SI workflow lets the analyst explore the full parameter landscape, observe how clusters emerge, split, merge, and dissolve, and progressively build knowledge about the spatial structure of the data---knowledge that no single run can provide and that constitutes analytical value independent of any particular parameter choice.

\textbf{Phase~1: Initial parameter sweep.} The iteration metrics chart displays per-iteration quality metrics as multi-line plots over the parameter axis~(T1), with warm colors for higher-is-better and cool colors for lower-is-better metrics. Iterations where all archetypes are represented are highlighted with dark gridlines and asterisks~(T8).

\begin{figure}[tb]
  \centering
  \begin{minipage}[c]{0.34\columnwidth}
    \begin{subfigure}{\linewidth}
      \centering
      \includegraphics[width=\linewidth,keepaspectratio]{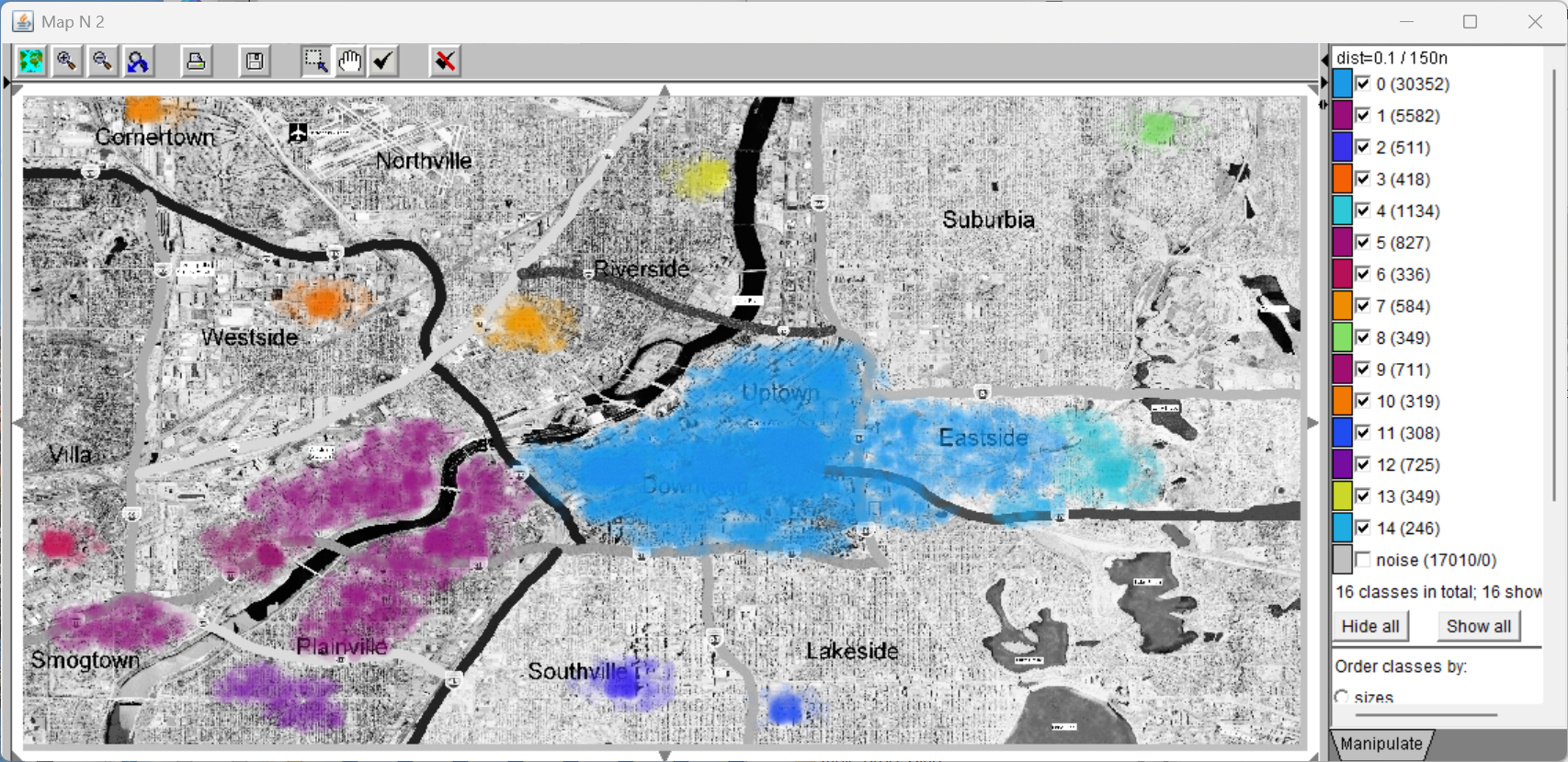}
    \end{subfigure}
    \begin{subfigure}{\linewidth}
      \centering
      \includegraphics[width=\linewidth,keepaspectratio]{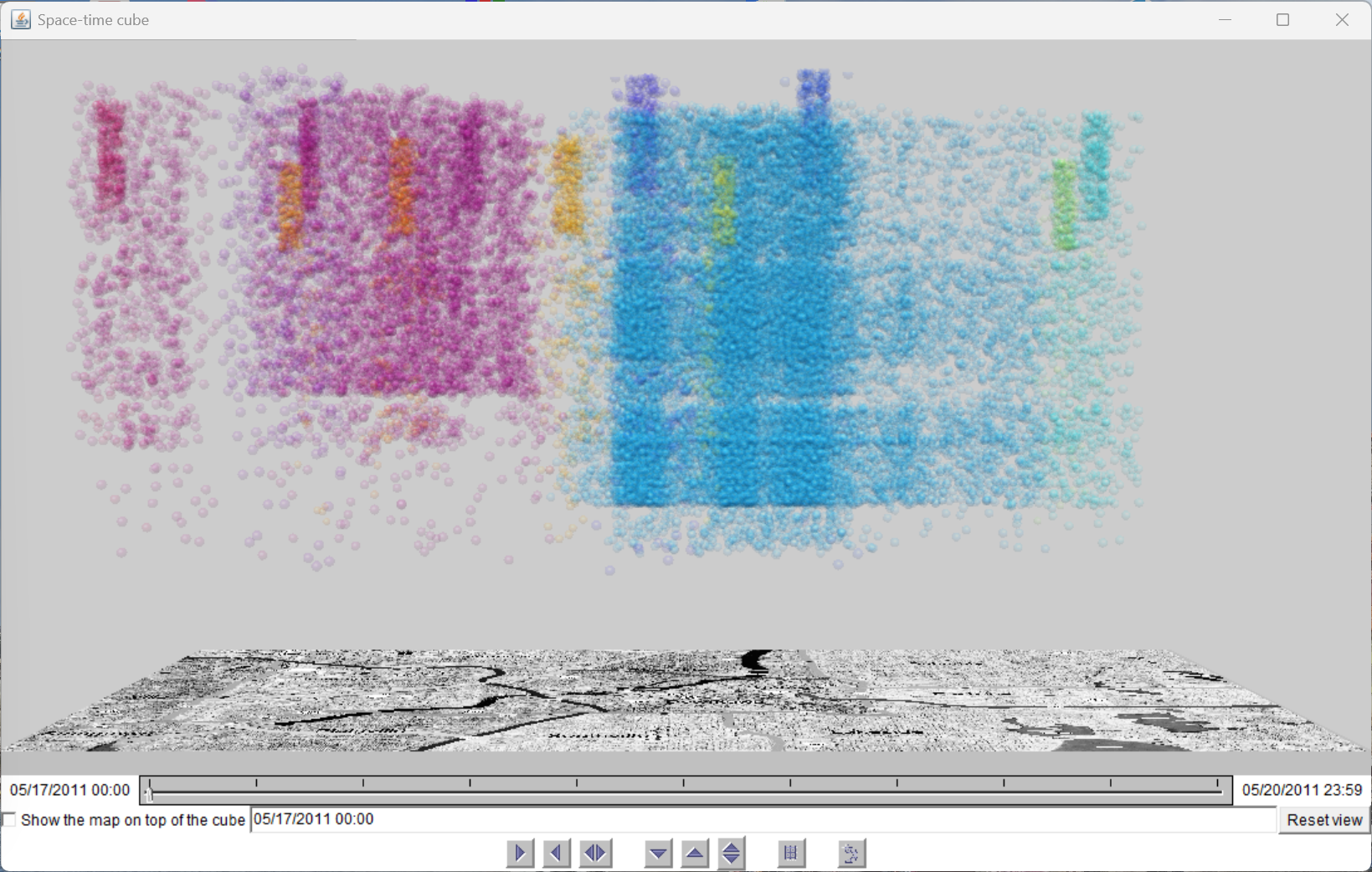}
      \caption{Map \& Space-time cube}
      \label{fig:iter1map}
    \end{subfigure}
  \end{minipage}%
  \hfill
  \begin{minipage}[c]{0.64\columnwidth}
    \begin{subfigure}{\linewidth}
      \centering
      \includegraphics[width=\linewidth,keepaspectratio]{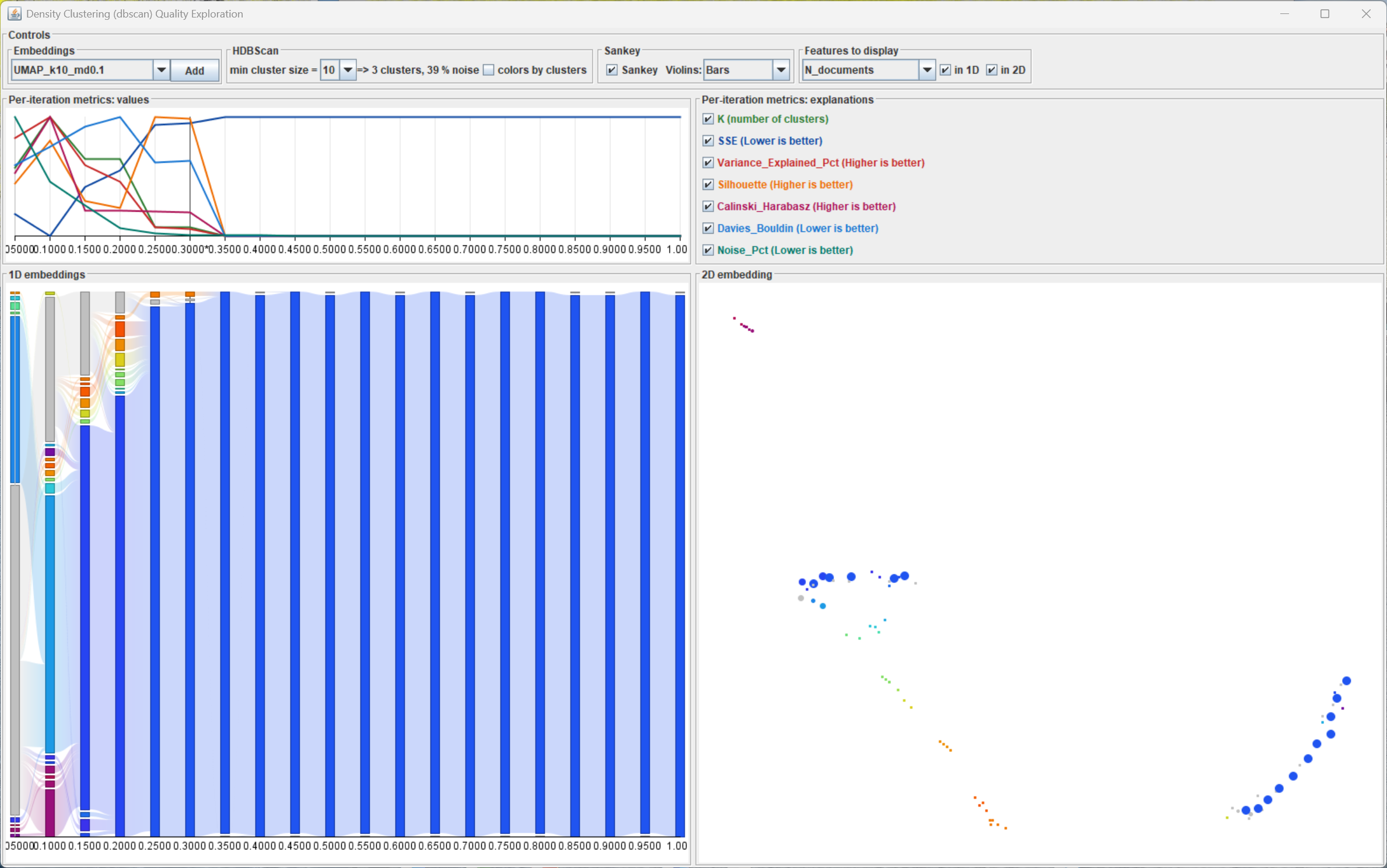}
      \caption{IteraScope}
      \label{fig:iter1is}
    \end{subfigure}
  \end{minipage}
  \vspace{-1mm}
  \caption{Round~1: $\varepsilon = 0.05 \ldots 1.0$, step~0.05, \texttt{min\_samples}~$=$~150. The metrics chart (top of~b) shows silhouette peaking near $\varepsilon = 0.1$; the Sankey view (bottom of~b) shows rapid cluster merging beyond $\varepsilon = 0.15$. The map and space-time cube~(a) reveal temporal layering of the contamination event. See Appendix~A, Section~A.2 for additional views.}
  \label{fig:iteration1}
\end{figure}

The analyst estimates that hospital clusters---the most conspicuous concentrations---comprise roughly 300~messages. Setting \texttt{min\_samples}~$=$~150 and sweeping $\varepsilon$ from 0.05 to 1.0 in steps of 0.05, the analyst launches the first computation round. (A preceding exploratory sweep with $\varepsilon$ up to~5.0 confirmed that useful structure exists only below $\varepsilon \approx 0.2$; see Appendix~A, Section~A.1.) The metrics chart (top of Fig.~\ref{fig:iter1is}) immediately reveals the coarse parameter landscape: at $\varepsilon = 0.05$ dozens of tiny clusters are dominated by noise; by $\varepsilon = 0.25$ nearly all messages collapse into one or two clusters. The intermediate value $\varepsilon = 0.1$ stands out: moderate $K$, silhouette near its peak, and manageable noise~\%.

\textbf{Phase~4: Domain confirmation.} (visited early to confirm domain relevance before detailed transition analysis)\textbf{.} Sharing the $\varepsilon = 0.1$ grouping to the linked map and space-time cube confirms this assessment and reveals temporal structure invisible in the metrics (Fig.~\ref{fig:iter1map}). On the map, clusters align with hospitals, the river corridor, and downwind neighborhoods. In the space-time cube, the first day shows only scattered messages; large clusters corresponding to wind- and water-dispersed contamination emerge on days~2--4; and tight hospital clusters appear on days~3--4 only, consistent with a delayed medical response. This temporal signature illustrates why domain-linked views are essential in Phase~4. (Additional views are in Appendix~A, Figures~A.6--A.9.)

\textbf{Phase~1 refined: Narrowing the distance threshold.}  The analyst narrows the range to $\varepsilon = 0.05 \ldots 0.15$ (step~0.01) and reruns the computation. The IS display (Appendix~A, Section~A.3) now provides finer resolution. 
Noise~\% remains substantial, but the discovered $K$ has stabilized and the Sankey transitions show that cluster structure at $\varepsilon = 0.07$ persists at $0.08$ and $0.09$ with only minor boundary adjustments. Several iterations around $\varepsilon = 0.07$ are marked complete, and in the 2D embedding these groups sit close to their archetype centroids. (Additional views are in Appendix~A, Figures~A.10--A.15.)
Sharing the $\varepsilon = 0.07$ grouping to the map confirms pattern validity: hospital clusters are tightly delineated, river-corridor clusters follow elongated geometries, and temporal layering is preserved with sharper boundaries.

\begin{figure}[!]
  \centering
  \begin{subfigure}{\columnwidth}
    \centering
    \includegraphics[width=.8\linewidth,keepaspectratio]{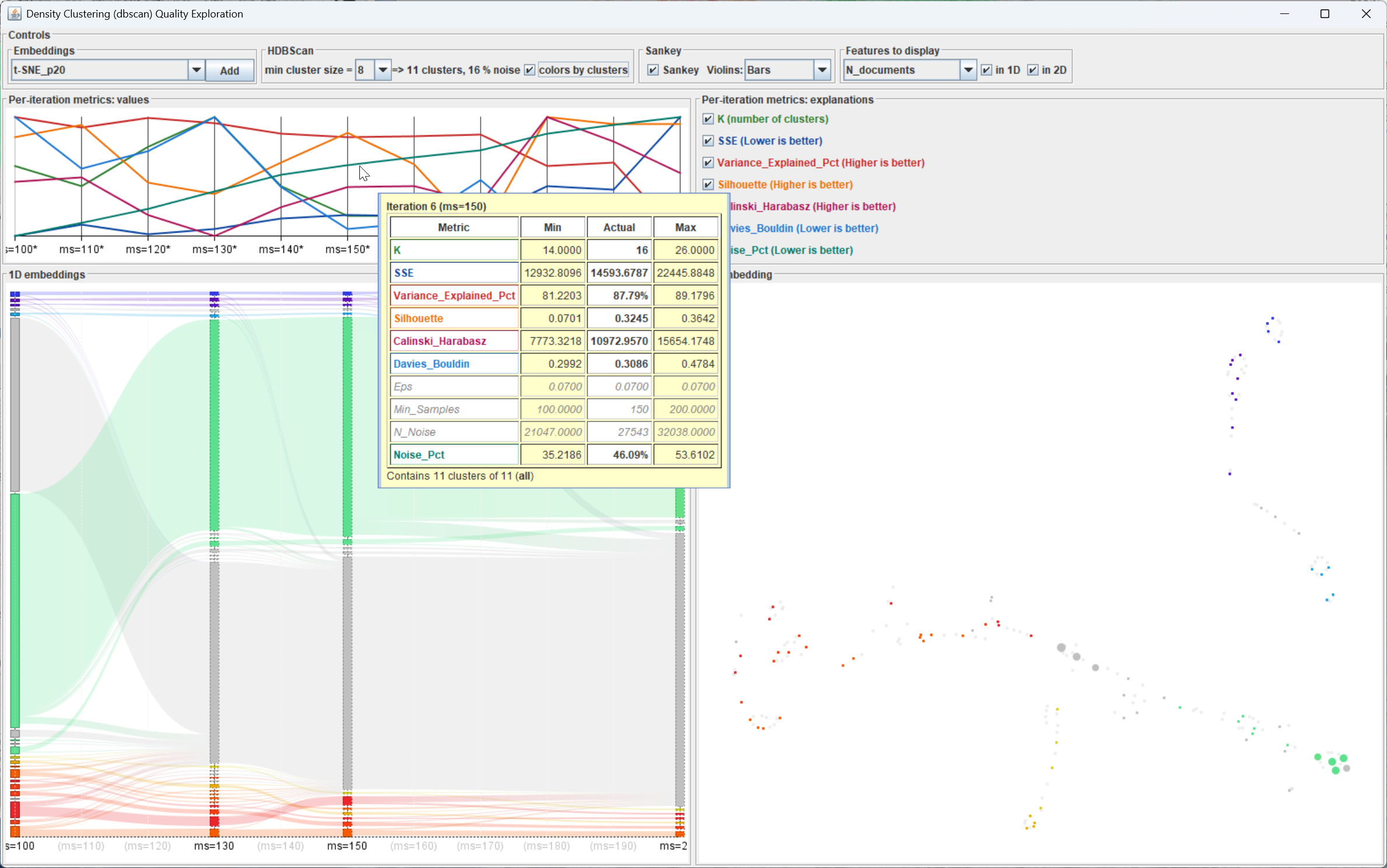}
    \caption{IS display with four visible iterations (100, 130, 150, 200).  Sankey bands between 130 and 150 reveal which clusters persist and which lose members to noise.}
    \label{fig:iter3_sankey}
  \end{subfigure}

  \vspace{2pt}

  \begin{subfigure}{\columnwidth}
    \centering
    \includegraphics[width=\linewidth,keepaspectratio]{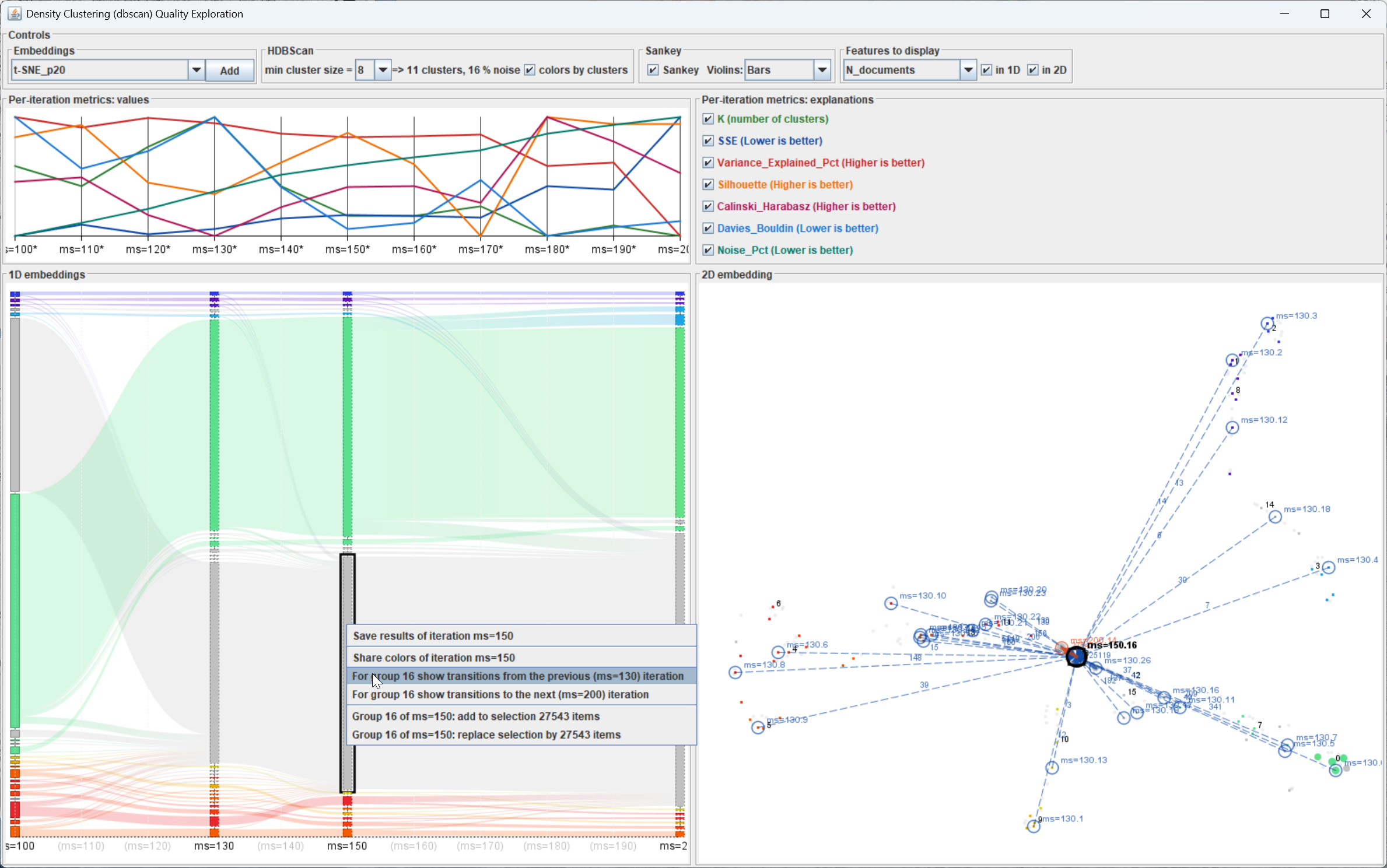}
    \caption{Transition from non-noise groups at \texttt{min\_samples}~$=$~130 to the noise group at 150.  The 2D embedding (right) shows connection lines with document counts.}
    \label{fig:iter3_transitions}
  \end{subfigure}

  \vspace{2pt}

  \begin{subfigure}{.8\columnwidth}
    \centering
    \includegraphics[width=\linewidth,keepaspectratio]{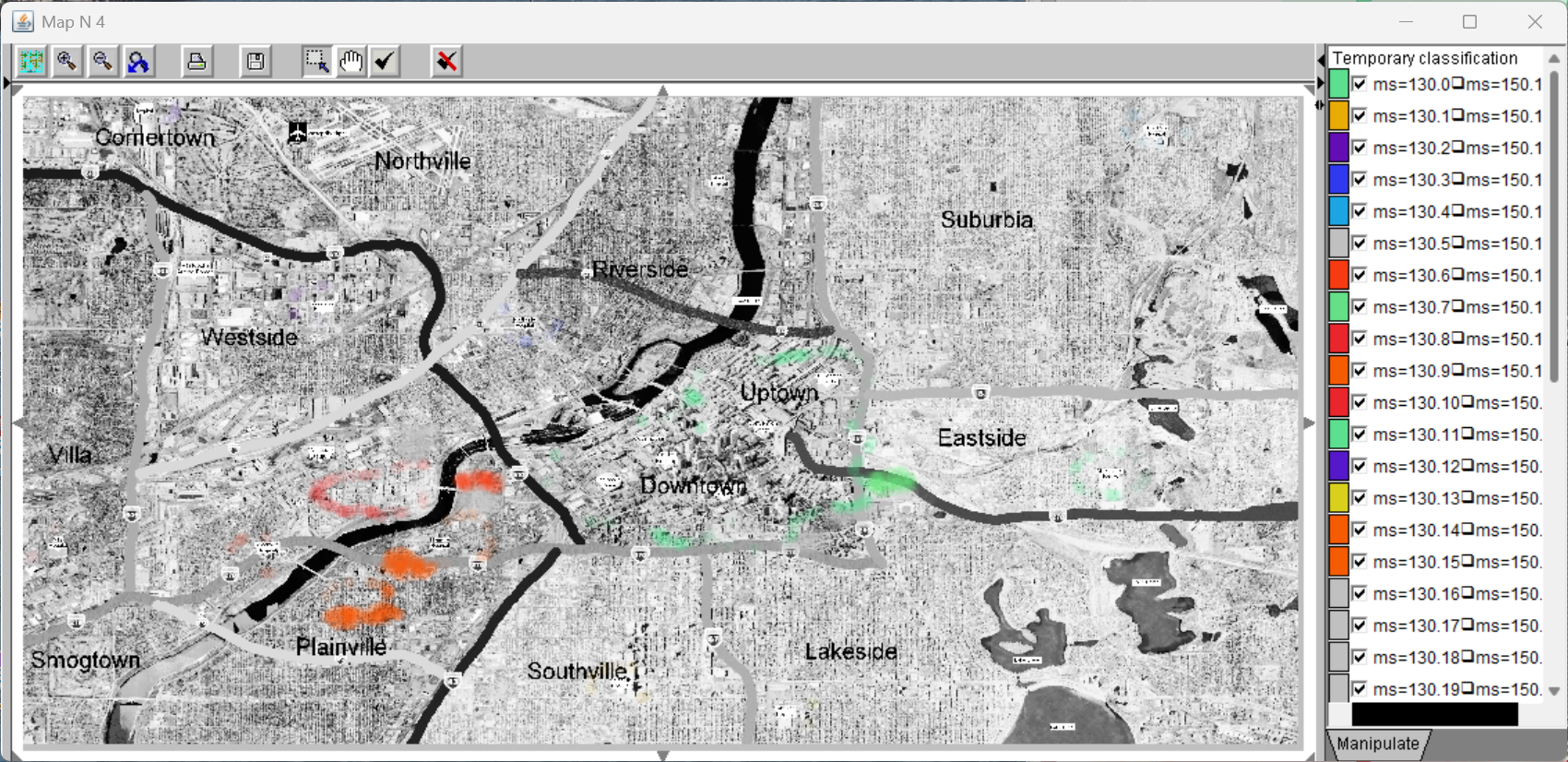}
    \caption{Map: borderline members are concentrated along cluster edges in the city center and river corridor.}
    \label{fig:iter3_map}
  \end{subfigure}

  \caption{Round~3: sweeping \texttt{min\_samples} from 100 to 200 (step~10, $\varepsilon = 0.07$). The analyst hides intermediate iterations, compares 130 and 150, and highlights messages transitioning to noise---they originate from cluster borders in the contamination-affected areas. See Appendix~A, Section~A.4.}
  \label{fig:iteration3}
\end{figure}

\textbf{Phases~2--3: Sweeping the neighborhood size.}  Holding $\varepsilon = 0.07$ fixed, the analyst sweeps \texttt{min\_samples} from 100 to 200 (step~10) to test robustness of the cluster structure to the density threshold (Fig.~\ref{fig:teaser}). Silhouette rises gently to about 150 and then plateaus, while noise~\% increases steadily. The analyst compares iterations 130 and 150 by hiding intermediate axes. The resulting Sankey view (Fig.~\ref{fig:iter3_sankey}) shows identical core structure: the same groups persist, differing only in noise boundaries. (The full sweep is documented in Appendix~A, Figures~A.16--A.19.)

\textbf{Phase~4: Transition inspection.}  To understand what is lost at the stricter threshold, the analyst selects \texttt{min\_samples}~$=$~150, clicks the noise group, and activates ``highlight transitions from'' to visualize messages that moved from non-noise clusters at 130 to noise at 150. This creates a temporary categorical attribute encoding origin and destination (e.g.\ ``130.3~$\to$~150.noise''), immediately available to all linked views for color-coding and filtering. In the 2D embedding (Fig.~\ref{fig:iter3_transitions}), connection lines show that several clusters each shed a modest number of borderline members. Propagating this transition class to the map (Fig.~\ref{fig:iter3_map}) reveals that these members are located at cluster edges---predominantly in the city center and along the river corridor---rather than at random locations. This geographic coherence confirms that the lost members are genuine peripheral observations; their small number and border positions indicate that \texttt{min\_samples}~$=$~150 provides cleaner separation without sacrificing meaningful structure. The analyst selects 150 as the final neighborhood-size parameter. (Additional views are in Appendix~A, Figures~A.20--A.24.)

Beyond this parameter choice, the exploration process yielded structural knowledge about the contamination event: the multi-scale density hierarchy (tight hospital clusters vs.\ diffuse dispersal patterns), the temporal layering (delayed medical response visible only in space-time), and the spatial coherence of boundary members---none of which could be obtained from any single clustering run.

\subsection{Partition-Based Clustering: EU Population Structure}
\label{sec:vis_partition}

The data comprise demographic indicators for ${\sim}1\,500$ European NUTS-3 regions from Eurostat's \texttt{demo\_r\_pjangrp3} table (the acquisition script is provided in Appendix~B). For each region, the dataset includes the percentage of female population and the distribution across age groups (0--14, 15--24, 25--44, 45--64, 65+). The analyst selects attributes reflecting gender and age structure as clustering input, excluding \texttt{pct\_male} and \texttt{pct\_65\_plus} to avoid linear redundancy.

Unlike the VAST scenario---where the challenge was choosing density parameters with an unknown number of clusters---the partition-based scenario poses a different question: \emph{how many clusters best capture Europe's demographic typology?} The SI workflow addresses this by sweeping~$K$ and using transition stability and domain context to judge when additional clusters cease to add interpretive value.

\begin{figure}[tb]
  \centering
  \includegraphics[width=\columnwidth,keepaspectratio]{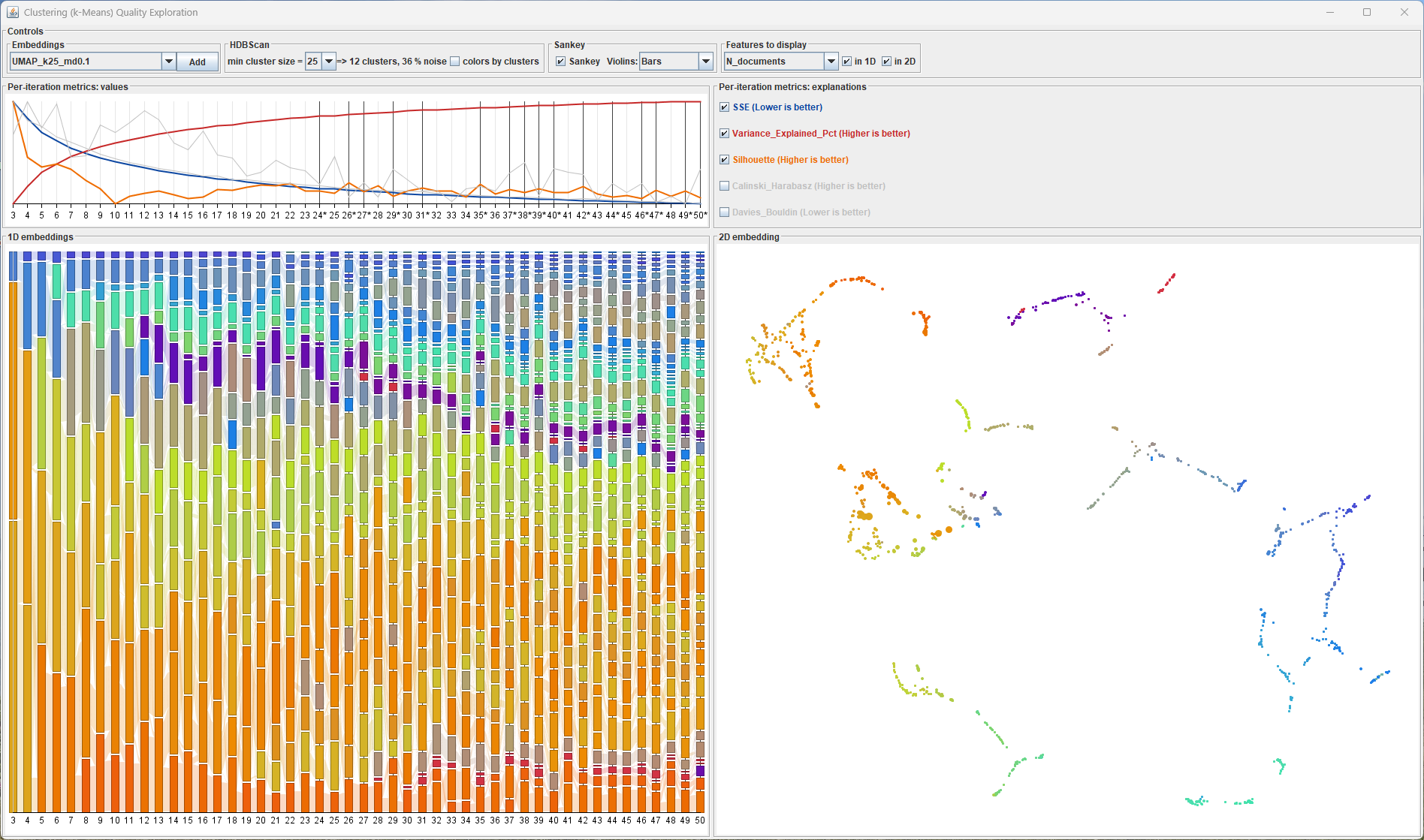}
  \caption{Partition-based clustering of EU NUTS-3 population data. 
  }
  \label{fig:pbc_quality}
\end{figure}

\textbf{Phase~1: Selecting the number of clusters.} The analyst runs K-means for $K = 3, \ldots, 50$ with a fixed seed and examines the IS metrics chart (Fig.~\ref{fig:pbc_quality}). SSE decreases monotonically with no sharp elbow, but silhouette peaks clearly at $K = 6$, followed by a secondary plateau from $K \approx 18$ to $\approx 24$. HDBSCAN archetype detection marks $K = 24$ as the first complete iteration; however, $K = 20$ achieves comparable silhouette and contains all but one archetype. The analyst decides to compare these two candidates. (Sankey transitions between $K = 20$ and $K = 24$ are documented in Appendix~B, Figures~B.2--B.3.)

\begin{figure}[tb]
  \centering
    \includegraphics[width=.6\linewidth,keepaspectratio]{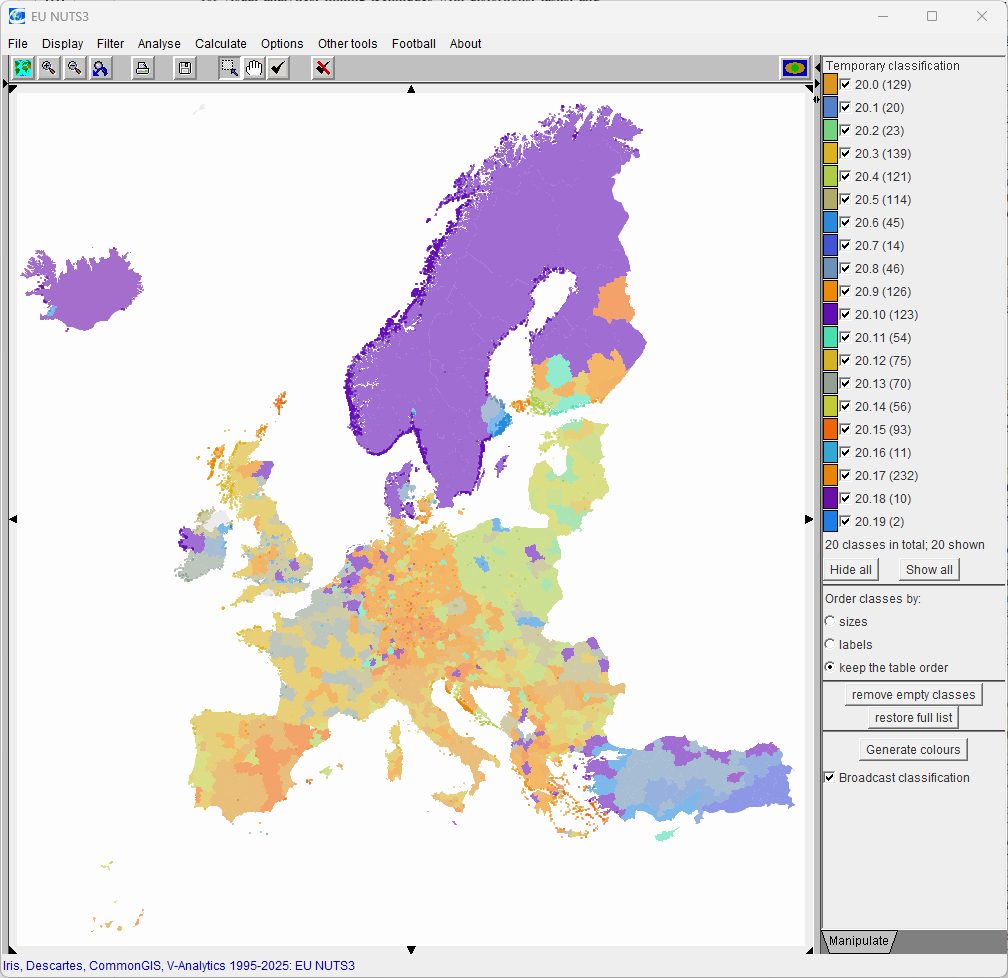}
    \includegraphics[width=.39\linewidth,keepaspectratio]{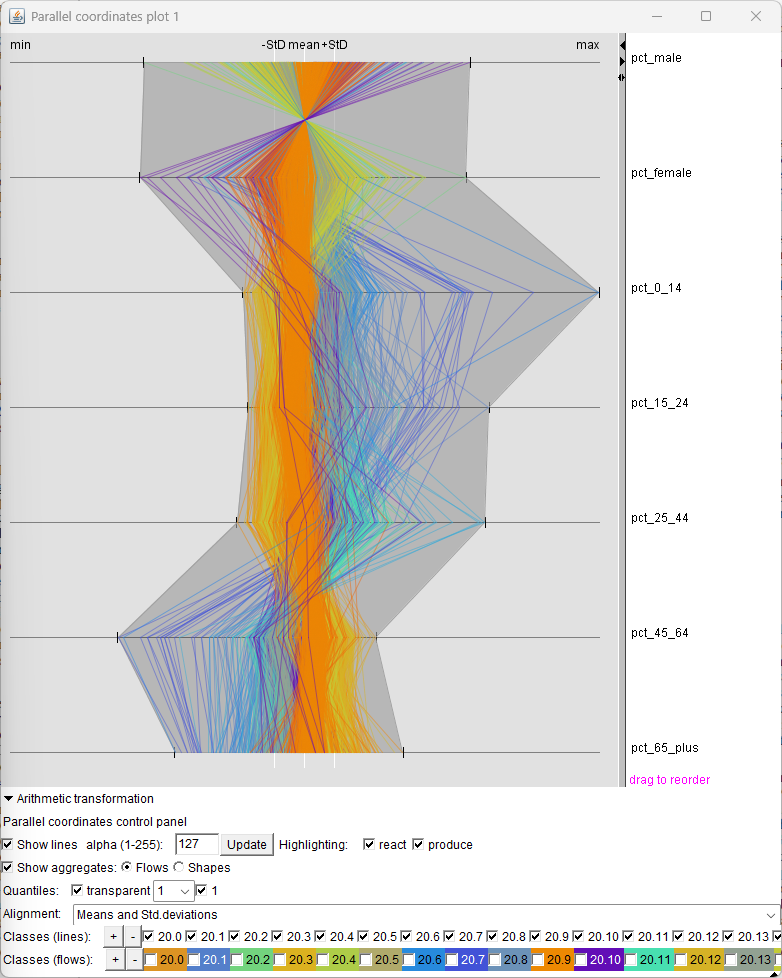}
    \caption{Left: Map for $K = 20$---the violet cluster (20.10) forms a spatially contiguous block. Right: Parallel coordinates---the violet cluster (highlighted) shows low shares of female and young age groups.}
    \label{fig:pbc_map20}
\end{figure}

\textbf{Phase~4: Comparing $K = 20$ and $K = 24$ in domain context.} (again visiting domain confirmation before detailed transition analysis to establish spatial context)\textbf{.}   Sharing both groupings to the map reveals similar macro-level spatial patterns---compact Nordic, Mediterranean, and Central-European groupings---but the $K = 24$ solution splits several spatially coherent regions without clear interpretive gain (see Appendix~B, Figures~B.6--B.7). A cluster of particular interest is the \emph{violet group} (labeled 20.10 at $K = 20$ and 24.7 at $K = 24$). At $K = 20$, this cluster forms a spatially contiguous block covering Scandinavia (Fig.~\ref{fig:pbc_map20}, left). At $K = 24$, the same demographic profile persists in cluster 24.7 but its spatial footprint fragments as border districts are reassigned to neighboring clusters. The parallel coordinates plot (Fig.~\ref{fig:pbc_map20}, right) reveals this group's characteristic profile. 

\begin{figure}[!]
  \centering
  \begin{subfigure}{\columnwidth}
    \centering
    \includegraphics[width=.8\linewidth,keepaspectratio]{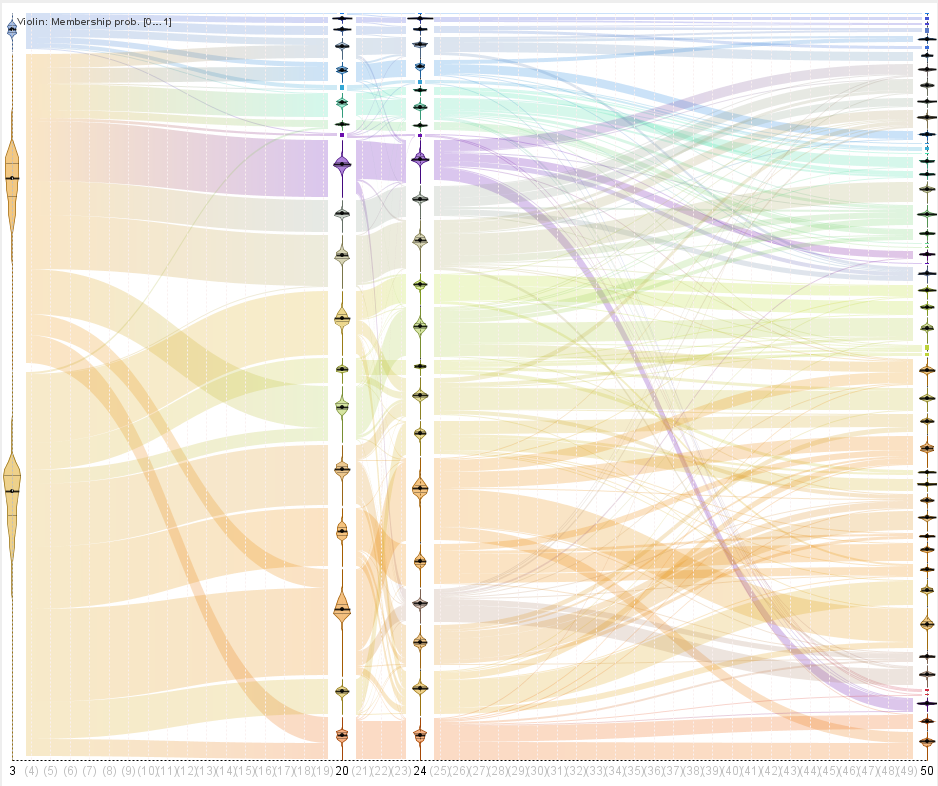}
    \caption{Violin plots of membership probability at $K = 20$ and $K = 24$.  Narrow violins near~1 indicate confident membership; wider lower tails signal borderline regions.}
    \label{fig:pbc_20_24}
  \end{subfigure}

  \vspace{2pt}

  \begin{subfigure}{\columnwidth}
    \centering
    \includegraphics[width=.6\linewidth,keepaspectratio]{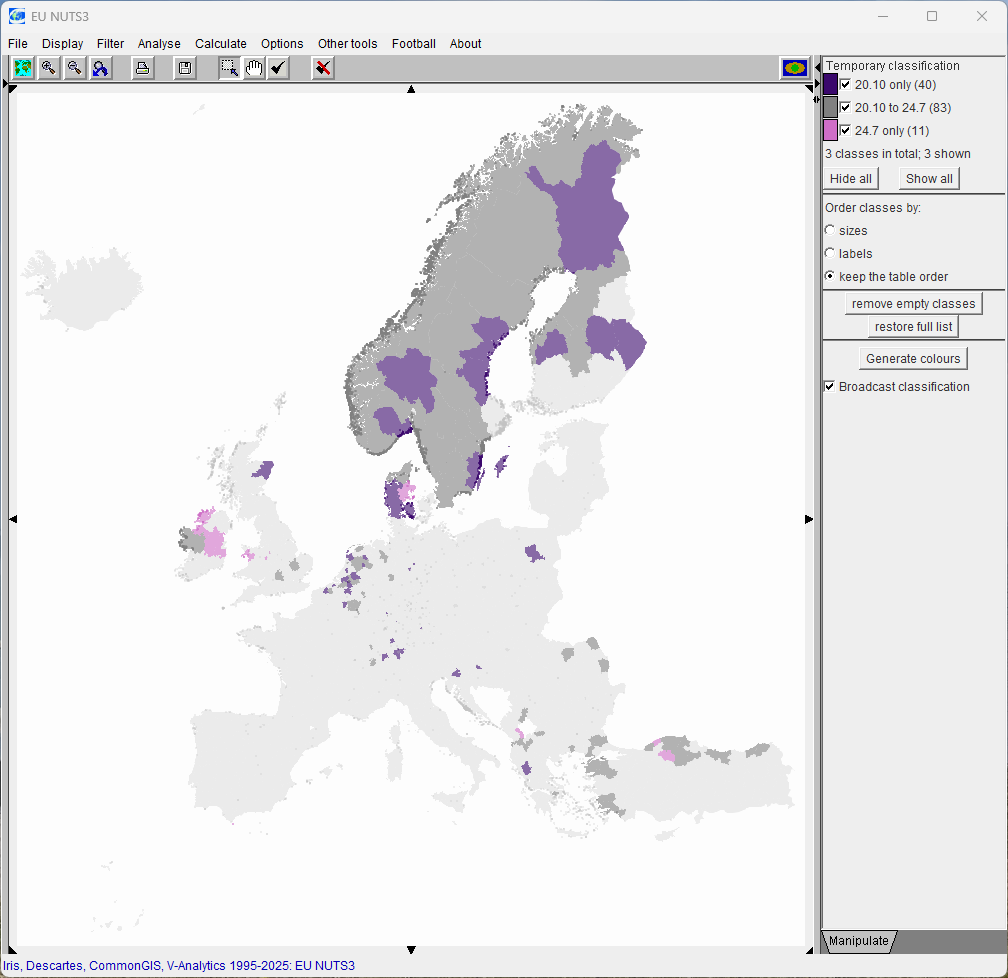}
    \includegraphics[width=.39\linewidth,keepaspectratio]{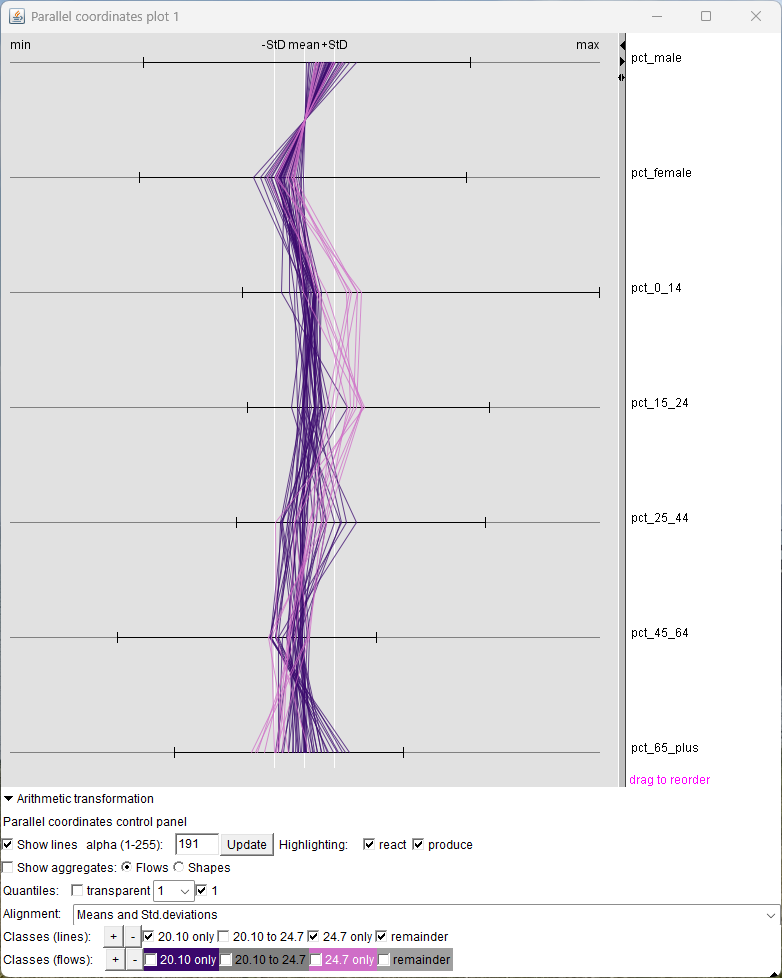}
    \caption{Transition 20.10\,$\to$\,24.7: members that persist in the violet cluster form a contiguous spatial core.}
    \label{fig:pbc_transition_specific}
  \end{subfigure}

  \vspace{2pt}

  \begin{subfigure}{\columnwidth}
    \centering
    \includegraphics[width=.6\linewidth,keepaspectratio]{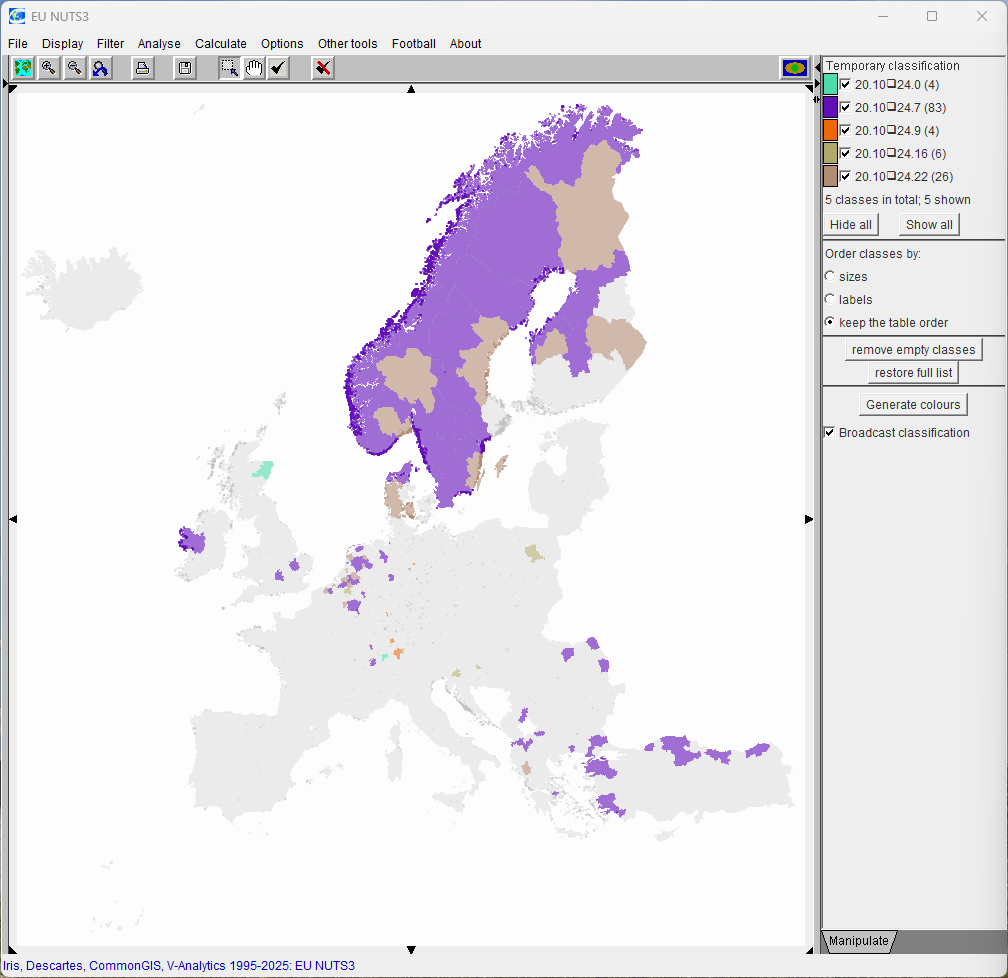}
    \includegraphics[width=.39\linewidth,keepaspectratio]{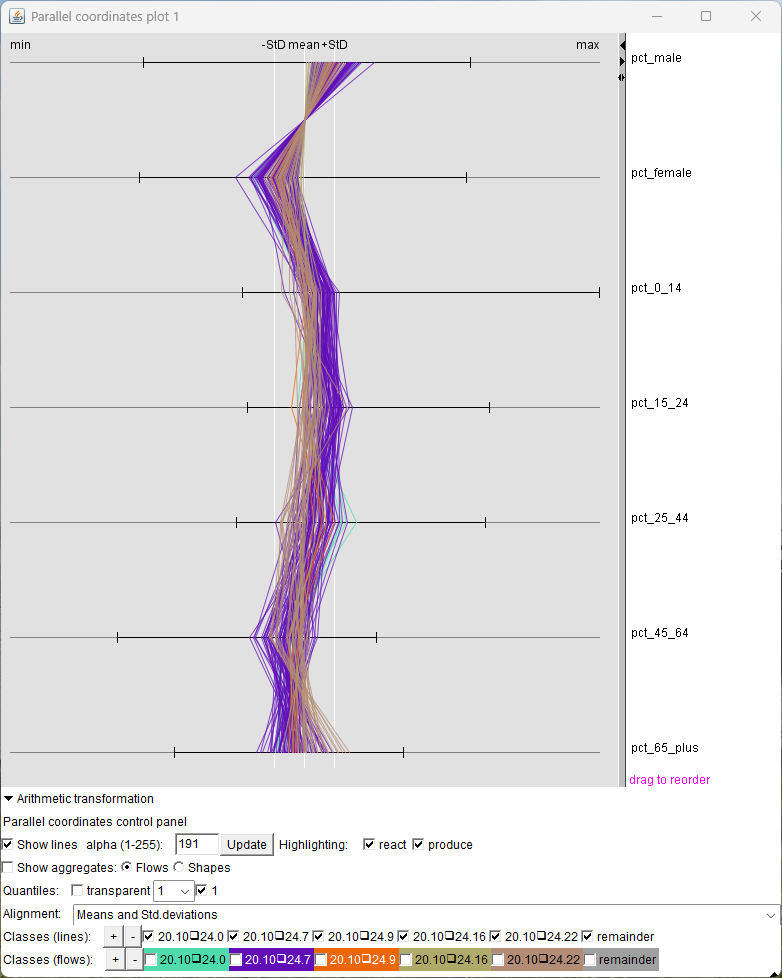}
    \caption{Transition 20.10\,$\to$\,24.*: all destinations color-coded.  Members lost from the violet cluster (non-violet colors) are scattered along its boundary.}
    \label{fig:pbc_transition_all}
  \end{subfigure}

  \caption{Transition and confidence analysis from cluster 20.10 ($K = 20$) to $K = 24$. Violin plots~(a) reveal within-cluster confidence at a glance. The specific transition to 24.7~(b) preserves the spatial core; the full decomposition~(c) shows that losses are peripheral and geographically dispersed.}
  \label{fig:pbc_transitions}
\end{figure}

\textbf{Phases~2--3: Transition and confidence analysis.}  The analyst activates violin plots for $K = 20$ and $K = 24$ to compare membership confidence~(T7). Fig.~\ref{fig:pbc_20_24} shows the resulting display. Violin shapes provide an immediate reading of assignment confidence: a narrow violin concentrated near the maximum indicates well-separated members, whereas a wide or bimodal violin signals borderline assignments. At $K = 20$, the violin for cluster~20.10 shows membership concentrated near~1 with a narrow tail. At $K = 24$, cluster~24.7 retains a similar median but develops a wider lower tail, reflecting the borderline districts that fragmented away. This contrast directly informs the analyst's judgment: $K = 20$ yields more internally coherent groupings, while the additional clusters at $K = 24$ absorb borderline regions without improving separation.

To examine the transition in detail, the analyst highlights the flow from 20.10 to 24.7 and propagates the resulting class attribute to the map (Fig.~\ref{fig:pbc_transition_specific}). The retained members form the cluster's spatial core---a contiguous mass across Scandinavia---confirming that the essential geographic structure survives the increase in~$K$. Next, highlighting the transition from 20.10 to \emph{all} clusters at $K = 24$ (creating a class attribute whose values encode each destination (``20.10$\,\to\,$24.7'', ``20.10$\,\to\,$24.3'', ``20.10$\,\to\,$24.11'', etc., see Fig.~\ref{fig:pbc_transition_all}) reveals that lost members scatter along the cluster boundary, reassigned to geographically adjacent clusters rather than forming a coherent sub-group. This confirms that the additional clusters merely shave off periphery without introducing new demographic archetypes.

\textbf{Phase~5: Seed stability and archetype verification.} To verify that the $K = 20$ result is not an artifact of a particular random initialization, the analyst reruns K-means with $K = 20$ across seeds 1--30. The IS display shows remarkable consistency: Sankey bands are nearly horizontal (no splits or merges), silhouette fluctuates by less than 0.01, and the 2D embedding shows tight archetype clusters with negligible drift (see Appendix~B, Figure~B.12). Seed~1 is selected because (i)~none of its 20~clusters falls into HDBSCAN noise and (ii)~the iteration is marked complete.

\textbf{Phase~6: Decision.} The analyst concludes that $K = 20$ with seed~1 provides a well-justified partition: it balances metric quality (silhouette plateau), spatial coherence (contiguous map patterns), demographic interpretability (distinct profiles in parallel coordinates), high membership confidence (concentrated violins), near-complete archetype coverage, and full robustness to random initialization. The violin plots played a decisive role: by revealing that $K = 20$ yields uniformly confident assignments while $K = 24$ introduces boundary ambiguity, they provided actionable visual evidence that no single scalar metric could convey.

The workflow also revealed domain knowledge independent of the parameter choice: Europe's demographic structure exhibits a clear center-periphery gradient, the northern aging pattern forms a geographically contiguous archetype, and boundary regions between typologies are spatially systematic rather than randomly distributed.

\subsection{Topic Modeling: IEEE VIS Papers}
\label{sec:vis_topics}

This demonstration introduces word clouds, term-level transition tooltips, and temporal prevalence analysis as IS components specific to text corpora.

\textbf{Data.} The IEEE VIS publication dataset~\cite{isenberg2017vispubdata} comprises ${\sim}3\,800$ papers (1990--2024) with titles, abstracts, keywords, and publication years. We concatenate title and abstract into a single document per paper. The NMF pipeline extracts the top~300 unigram terms and 200 bigrams from the resulting TF-IDF matrix.

\textbf{Phase~0: Data cleaning.}  An initial NMF sweep over $n_{\mathrm{topics}} = 5 \ldots 25$ reveals a large, stable topic whose top keywords include \texttt{etx}, \texttt{lt}, and \texttt{gt}---clearly artifacts. Inspection confirms that 283~papers contain the HTML entity \texttt{\&lt;\&lt;ETX\&gt;\&gt;} appended to their abstracts (a legacy encoding error). Removing these artifacts and recomputing TF-IDF eliminates the spurious topic. (See Appendix~C, Figure~C.1.)

\begin{figure}[tb]
  \centering
  \includegraphics[width=\columnwidth,keepaspectratio]{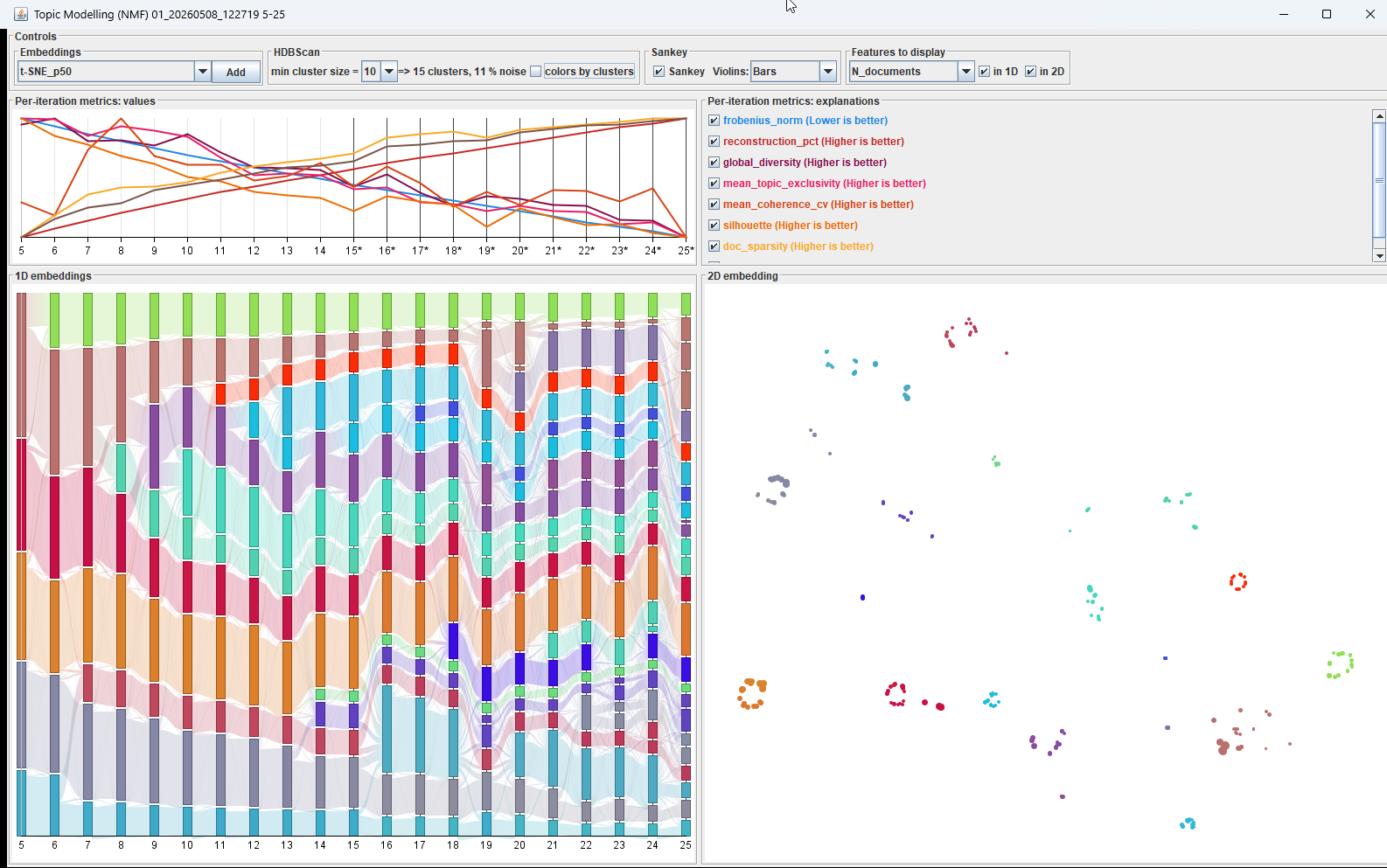}
  \caption{NMF topic modeling ($n_{\mathrm{topics}} = 5 \ldots 25$) after data cleaning. The bar chart shows topic counts per iteration with color-coded archetype membership; the 2D embedding (right) confirms tight archetype clusters. Only ${\sim}11\%$ of topic instances are HDBSCAN noise. See Appendix~C, Section~C.2.}
  \label{fig:tm_stability}
\end{figure}

\textbf{Phase~1: Parameter sweep and stability.}  After cleaning, the analyst reruns NMF for $n_{\mathrm{topics}} = 5 \ldots 25$. The IS display (Fig.~\ref{fig:tm_stability}) shows good overall stability: approximately 15~recurrent thematic archetypes emerge, and HDBSCAN marks only ${\sim}11\%$ of topic instances as noise. Reconstruction error decreases smoothly without a sharp elbow; coherence stabilizes in the range $n_{\mathrm{topics}} = 14 \ldots 18$. The 2D embedding confirms tight archetype clusters. The analyst identifies $n_{\mathrm{topics}} = 15$ and $16$ as candidates.

\begin{figure}[tb]
  \centering
  \includegraphics[width=\columnwidth,keepaspectratio]{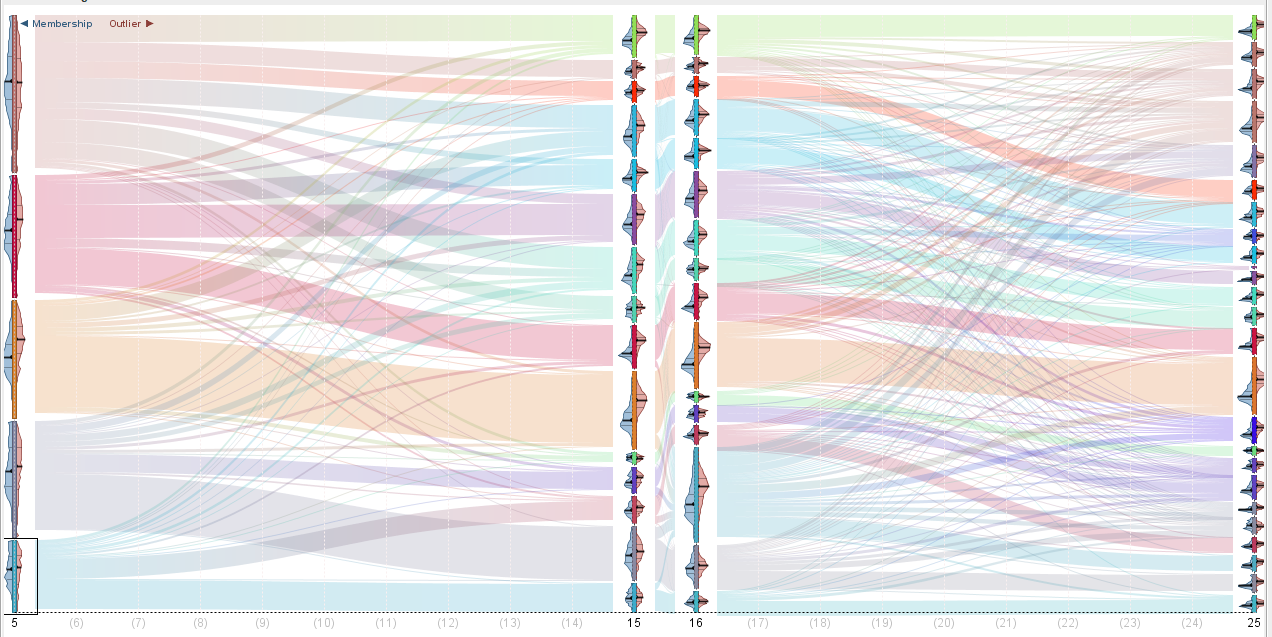}
  \caption{Sankey transitions between $n_{\mathrm{topics}} = 15$ and $16$ with violin plots in split mode. The new ``method'' topic (cyan) draws thin bands from multiple parents---a wastebasket pattern. Violins confirm higher membership confidence at 15~topics. See Appendix~C, Section~C.2.}
  \label{fig:tm_sankey}
\end{figure}

\textbf{Phases~2--3: Comparing 15 and 16 topics.} The analyst activates Sankey transitions between $n_{\mathrm{topics}} = 15$ and $16$ with violin plots in split mode (Fig.~\ref{fig:tm_sankey}). At $n_{\mathrm{topics}} = 16$, a new topic labeled ``method'' (cyan) appears with 662~documents. The Sankey bands reveal that this topic does not emerge from a clean split of a single parent; rather, it collects method-related terms from \emph{multiple} topics---visual analytics, graphs, volume rendering---each losing a thin band. This aggregation pattern signals a ``wastebasket'' topic that conflates methodological vocabulary across unrelated areas rather than capturing a genuine thematic community.

\begin{figure}[tb]
  \centering
  \includegraphics[width=.49\columnwidth,keepaspectratio]{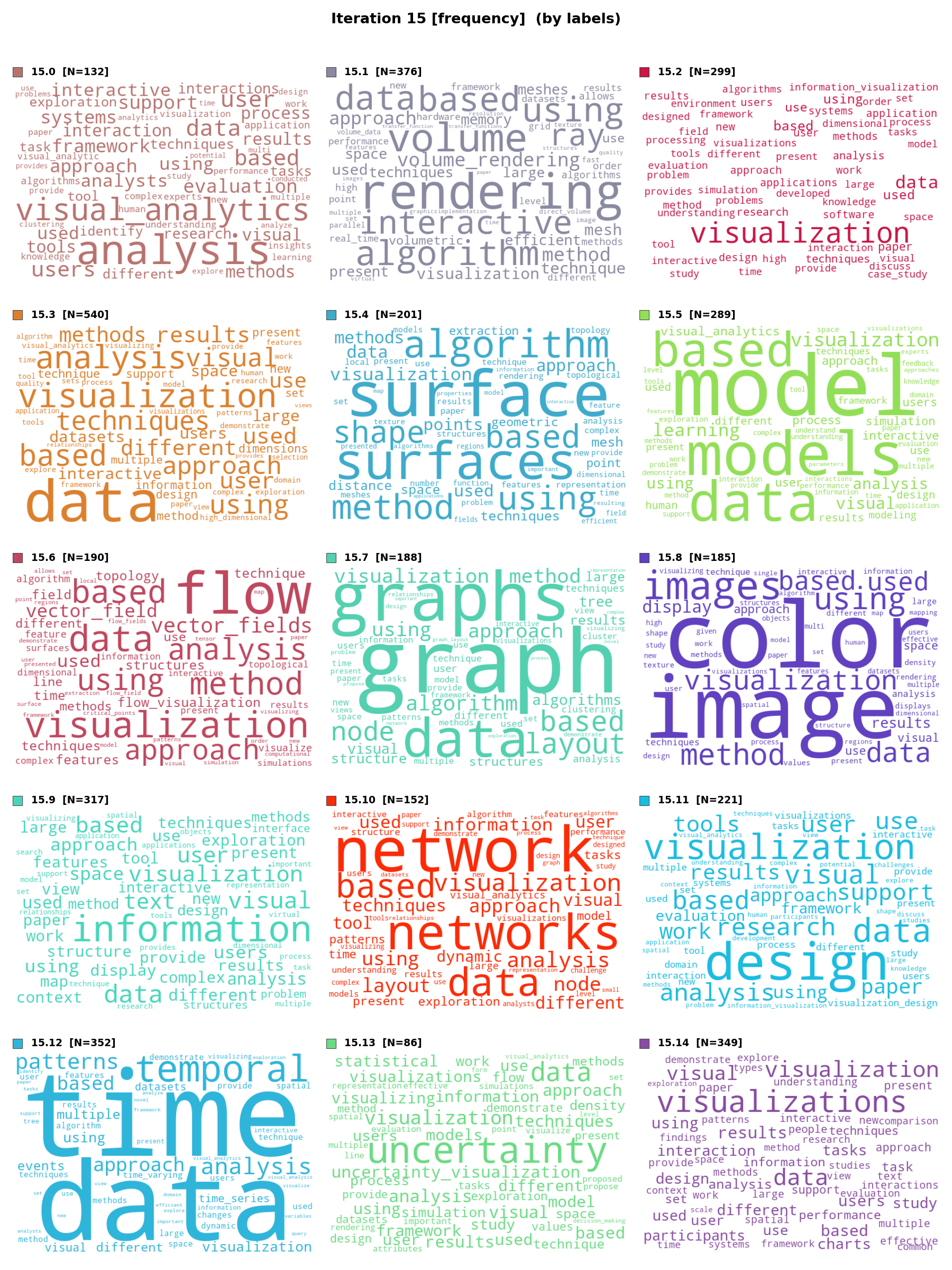}
  \includegraphics[width=.49\columnwidth,keepaspectratio]{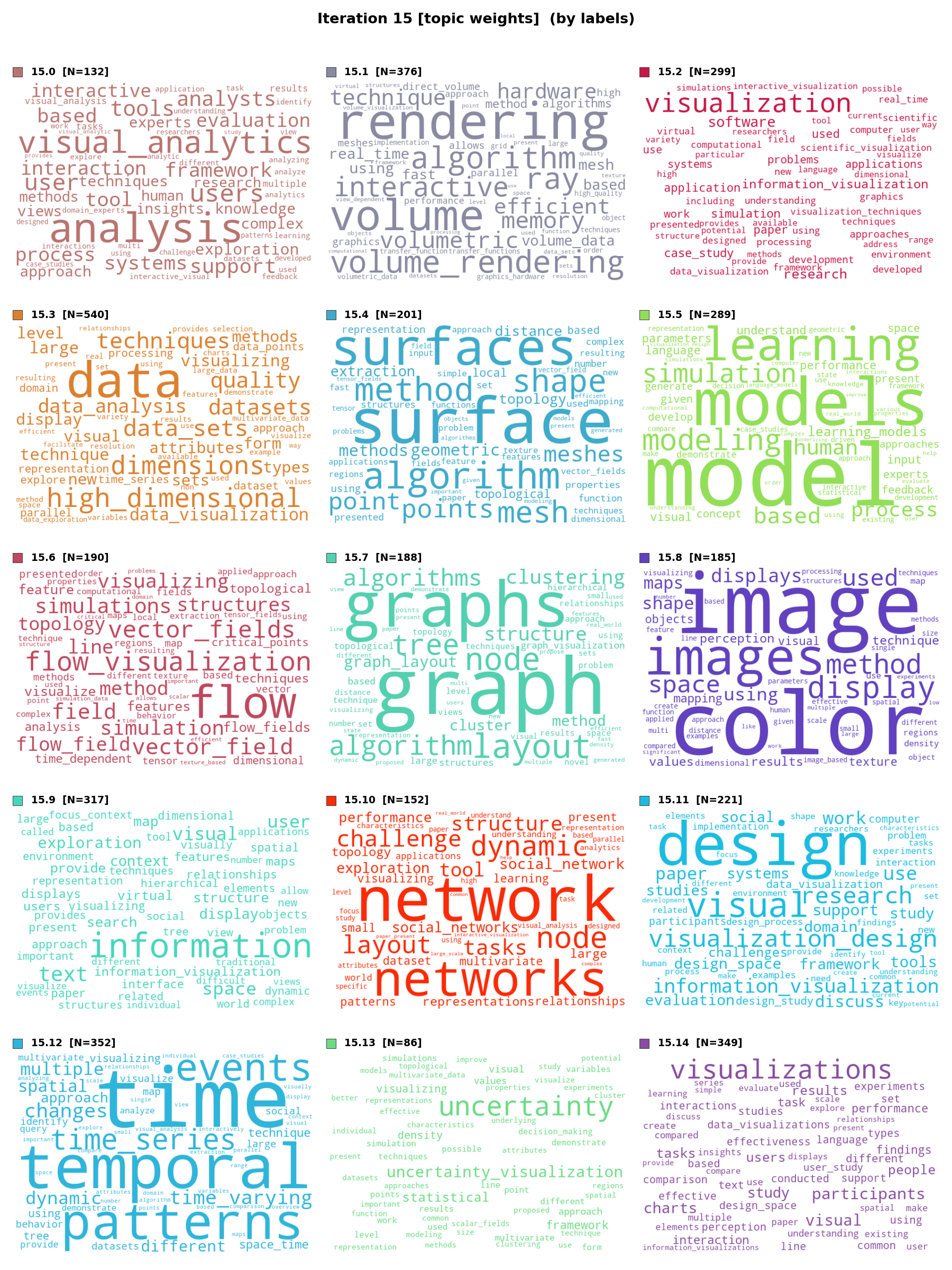}
  \caption{Word clouds (frequency-weighted on the left, and term-weighted on the right) for the selected 15-topic solution.  Each cell shows the top terms for one topic.  Thematic profiles are clearly distinct.}
  \label{fig:tm_wc15}
\end{figure}

Word clouds confirm this diagnosis: at $n_{\mathrm{topics}} = 15$, each topic exhibits a coherent thematic profile (Fig.~\ref{fig:tm_wc15}), whereas the additional topic at $16$ mixes generic terms (\texttt{method}, \texttt{approach}, \texttt{technique}, \texttt{based}, \texttt{model}) without a unifying theme (see Appendix~C, Section~C.2). Violin plots at $n_{\mathrm{topics}} = 15$ show narrow distributions concentrated near the maximum, indicating confident assignments. The analyst selects $n_{\mathrm{topics}} = 15$.

\begin{figure}[tb]
  \centering
  \includegraphics[width=\columnwidth]{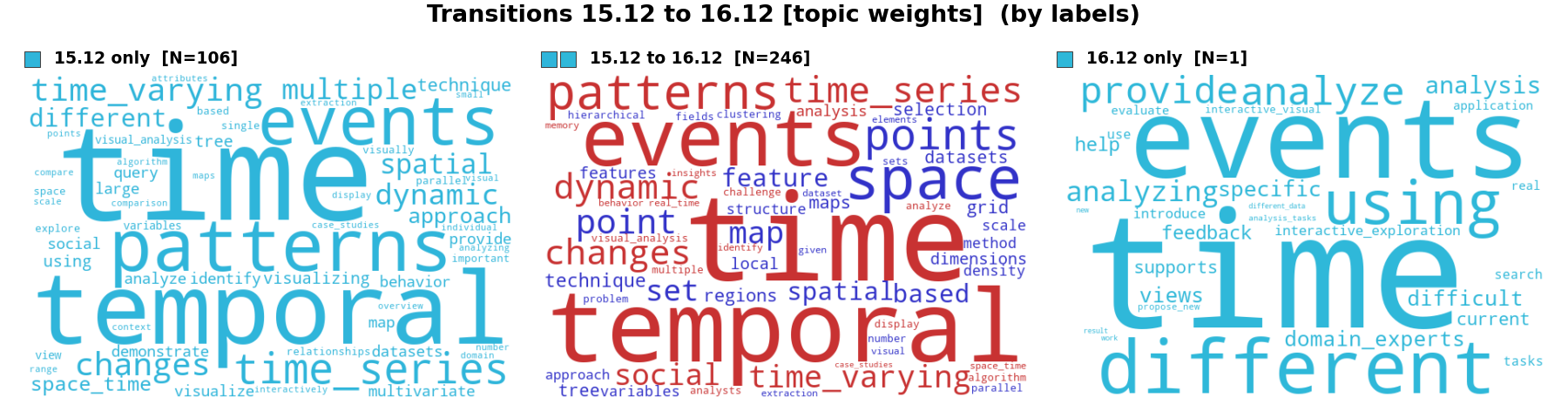}
  \caption{Term-weight word clouds for topic~12 across the 15-to-16 transition. Left: term weights in topic~15.12. Right: term weights in topic~16.12. Center: weight changes from 15.12 to 16.12---size encodes magnitude, color encodes sign (blue = decrease, red = increase). The lost terms (blue, center) are dominated by spatio-temporal vocabulary, confirming that these papers migrate to other topics when a 16th component is introduced.}
  \label{fig:time_topic_lost_tw}
\end{figure}

\textbf{Phase~4: Investigating the ``time'' topic.} Topic~12 relates to temporal data (top terms: \texttt{time}, \texttt{temporal}, \texttt{series}, \texttt{event}, \texttt{sequence}). It shrinks substantially at $n_{\mathrm{topics}} = 16$. The Sankey view (Appendix~C, Fig.~C.9) reveals that spatio-temporal papers migrate to the spatial topic, while specialized aspects---time-varying graphs, temporal event sequences, streaming data---wash away to their corresponding thematic topics. Of the 352~documents in topic~15.12, 246 persist in 16.12; the term-weight word cloud for the topic~15.12 (Fig.~\ref{fig:time_topic_lost_tw}) includes \emph{spatial} and \emph{space-time}---precisely the vocabulary that migrates once a 16th component becomes available. Examining the broader transition from $n_{\mathrm{topics}} = 5$ to $25$ confirms that the time topic progressively differentiates: at 15 it encompasses all temporal research; at higher~$K$ its subtopics disperse. This validates 15 as the granularity at which temporal research remains a coherent community.

\begin{figure}[tb]
  \centering
  \includegraphics[width=\columnwidth,keepaspectratio]{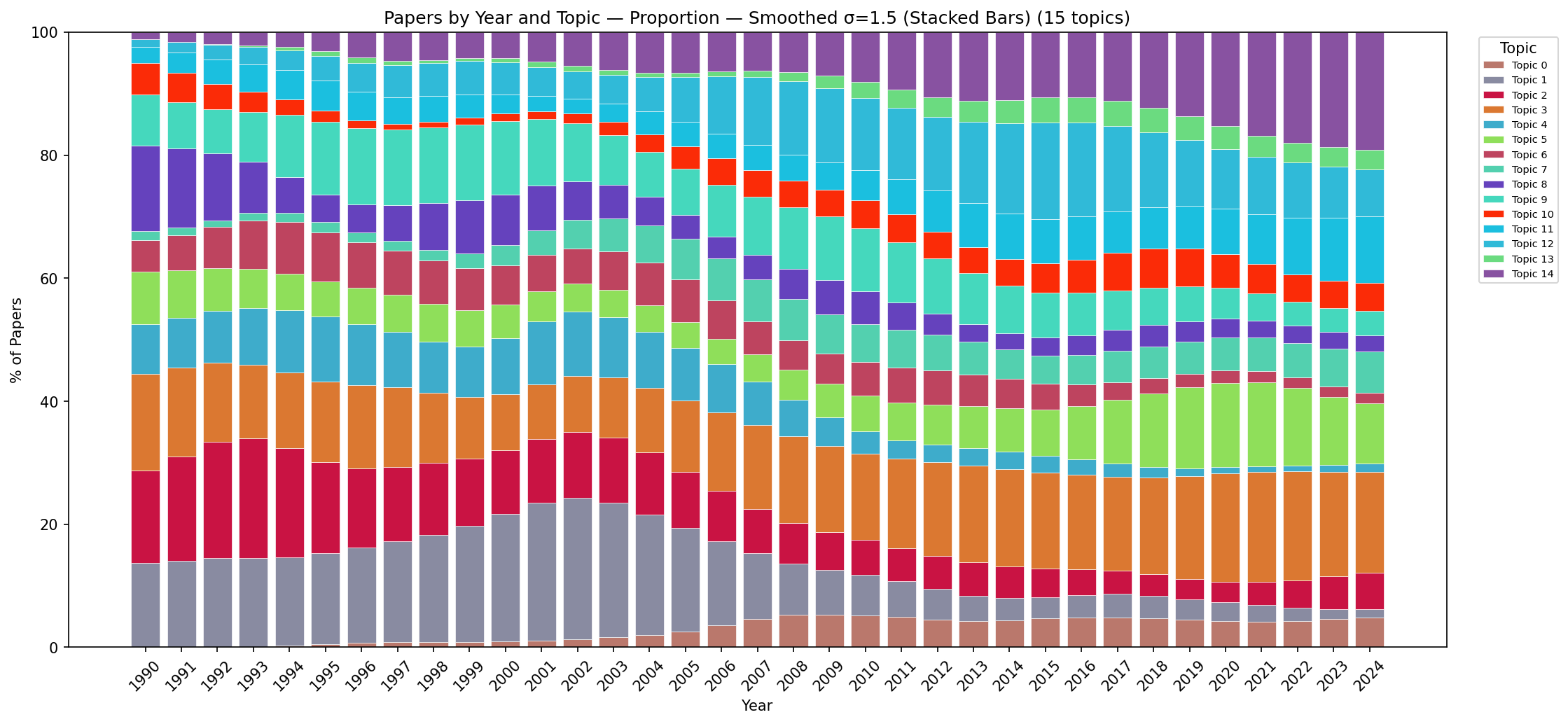}
  \caption{Temporal prevalence of the 15~topics (proportional stacked areas, Gaussian-smoothed). See Appendix~C, Section~C.4 for alternative visualizations (line charts, absolute counts, raw data).}
  \label{fig:tm_temporal}
\end{figure}

\textbf{Phase~4 continued: Temporal prevalence analysis.} To interpret the 15-topic solution in the context of the VIS community's evolution, the analyst aggregates papers by dominant topic and publication year. Fig.~\ref{fig:tm_temporal} shows proportional stacked areas with Gaussian smoothing, revealing clear long-term trends:

\begin{itemize}
  \item \emph{Visual analytics} (topic~0) emerged around 2004 and stabilized at ${\sim}15\%$ share, coinciding with the VAST conference track.
  \item \emph{High-dimensional data} (topic~3) maintains a constant ${\sim}8$--$10\%$ share across the entire 33-year span.
  \item \emph{User studies} (topic~14) shows sustained growth from ${<}5\%$ before 2005 to ${\sim}20\%$ by 2024.
  \item \emph{Volume rendering} (topic~1) declines from dominant shares in the 1990s to below $5\%$, as the community broadened toward information visualization and analytics.
\end{itemize}

\noindent These trends align with known shifts in the VIS landscape and confirm that the discovered topics capture genuine, temporally coherent research communities.

\textbf{Phase~6: Decision.} The analyst concludes that NMF with $n_{\mathrm{topics}} = 15$ is well justified: it balances stability (${\sim}89\%$ non-noise archetypes), thematic coherence (distinct word-cloud profiles), membership confidence (concentrated violins), and temporal validity (prevalences align with known community evolution). The 16-topic alternative was rejected because its additional topic acts as a methodological wastebasket.

The exploration also generated knowledge about the VIS community itself: temporal research serves as an integrative hub that fragments along domain lines at finer granularity, spatio-temporal papers form a bridge population with dual thematic allegiance, and the community's research priorities have undergone three major shifts over 34 years---insights that emerge only from studying how topics differentiate across configurations.

\subsection{IteraScope: Components and Interactions}
\label{sec:vis_summary}

The three demonstrations introduced IS components progressively, each building on the previous. We now provide a consolidated reference of all components and their interactions, independent of any specific dataset or method family.



Three recurring design principles underpin IS: \emph{coordinated multi-view linking} (any operation is reflected in all views), \emph{progressive disclosure} (iterations can be hidden, violin modes toggled, word-cloud weights switched), and \emph{archetype-driven coloring} (consistent colors across iterations regardless of internal numbering).

\subsubsection{1D Embedding with Sankey Transitions}
\label{sec:vis_1d}

Each iteration occupies a vertical axis; groups are positioned by the 1D projection. In \textbf{Sankey mode}, bars whose height is proportional to group size are connected by gradient-colored bands whose width encodes the shared-document count~(T2). 
\textbf{Axis visibility} is toggled by clicking labels; hidden axes appear as dashed lines and Sankey neighbors are recalculated among visible axes only.

\textbf{Hover interactions.} Hovering over a column highlights its groups in the 2D embedding; hovering over a Sankey band highlights the connector as a dotted line with \texttt{iter.topic} labels. For topic modeling, group tooltips show top-$N$ terms with weights alongside per-group metrics. Connector tooltips show four aligned columns: \emph{From} terms, \emph{Lost} ($\Delta < 0$), \emph{Gained} ($\Delta > 0$), and \emph{To} terms---enabling fine-grained inspection of vocabulary changes during splits or merges.

\textbf{Click interaction.} Left-clicking a group shows its connections to the previous and next visible iterations in 2D: incoming in navy-blue dashes, outgoing in orange dash-dot, with line width proportional to shared-document count and arrowheads indicating direction. Clicking the same group again deactivates the selection.

\textbf{Context menu.} Right-click offers: \emph{Share colors} (propagating the full grouping to linked views), \emph{Show transitions from/to} (creating a class attribute with predecessor/successor origins), \emph{Show transition details} (three-class decomposition: ``left only'', ``shared'', ``right only''), and \emph{Select/Highlight} group members in the data table. Each sharing action optionally triggers word-cloud generation. 

This component served as the primary stability-assessment tool in all three demonstrations (Figs.~\ref{fig:iter1is},~\ref{fig:pbc_transitions},~\ref{fig:tm_sankey}).

\subsubsection{Violin Plots}
\label{sec:vis_violin}

When membership or outlier data is available~(T7), each Sankey bar can be replaced by a violin showing the kernel density estimate of membership probability, outlier score, or both in a split layout (membership on the left in blue, outlier on the right in red). Bandwidth is computed globally (Silverman's rule across all non-noise groups), and the maximum probability density is normalized globally so that violin widths are \emph{directly comparable} across all groups and iterations. Stat lines mark the median (thick), Q1, and Q3 (thin). Noise groups are rendered as plain gray bars.

The violin display proved decisive in Sections~\ref{sec:vis_partition} (Fig.~6a) and~\ref{sec:vis_topics} (Fig.~8).

\subsubsection{2D Group Embedding}
\label{sec:vis_2d}

All groups from all iterations are projected into a shared 2D space via the user-selected projection on group-level feature vectors (centroids for clustering; topic-weight vectors for NMF). Five rendering passes ensure visual clarity: (1)~hidden iterations as faded dots; (2)~visible non-highlighted items at full size; (3)~highlighted column on top with labels; (4)~hovered Sankey connector as a dotted line; (5)~clicked-group connection web with arrowheads and shared-document counts~(T3). Dot color is determined by HDBSCAN archetype assignment; the user toggles between per-item color (individual 2D position) and per-archetype color (archetype centroid position).

The 2D embedding confirmed archetype proximity (Section~\ref{sec:vis_density}), seed robustness (Section~\ref{sec:vis_partition}), and global stability (Section~\ref{sec:vis_topics}).

\subsubsection{Word Clouds and Term Inspection}
\label{sec:vis_wordclouds}

For topic modeling, the system generates word clouds per class label using two weighting modes: \emph{term frequency} (standard tf in the class's documents) and \emph{topic weights} (NMF $H$-matrix values). Transition classes use weight \emph{differences}: gained terms appear in green, lost terms in red. Clouds can be ordered by labels, by 1D-embedding similarity, or by document count.

The dual-weighting design allows the analyst to distinguish common vocabulary from truly discriminative terms (Figs.~\ref{fig:tm_wc15},~\ref{fig:time_topic_lost_tw}; Appendix~C, Figs.~C.5--C.6).

\subsubsection{Domain-Linked Views}
\label{sec:vis_domain}

When a grouping is \emph{shared} (via context menu), it is stored as a categorical class attribute in the data table. All linked displays---maps, parallel coordinates, space-time cubes, temporal prevalence charts---use class colors for rendering and support class-based filtering and aggregation~(T5, T6). Sharing modes include: \emph{Color} (full iteration), \emph{ColorTo/ColorFrom} (predecessors/successors of a group), and \emph{ColorShowTransitions} (three-class decomposition of a connector). Each mode triggers word-cloud generation if the text modality is active.

The demonstrations exercised domain views for geographic validation (Fig.~2a), multivariate profiling (Fig.~5), and temporal prevalence checking (Fig.~11).

\section{Parameter-Supervision Workflows}
\label{sec:workflows}

This section presents the paper's main contribution: three structured workflows---one per method family---that guide an analyst through systematic exploration of grouping results, building progressively deeper understanding of data structure and ultimately supporting an informed parameter decision. The workflows share a common six-phase skeleton (Section~\ref{sec:wf_common}) but differ in method-specific indicators. Table~\ref{tab:workflows} provides a consolidated reference mapping each phase to its key indicators and supporting IS components. 
Section~\ref{sec:wf_synthesis} distills the cross-workflow principles that constitute the deeper methodological contribution: the structural commonalities, reasoning patterns, and diagnostic mechanisms that emerge from treating parameter supervision as a first-class analytical activity.

\begin{table*}[tb]
  \centering
  \caption{Consolidated workflow reference. Each cell provides general guidance for the phase--method combination, followed by key indicators to monitor. Minimal back-references to Section~\ref{sec:visual} demonstrations are given in brackets.}
  \label{tab:workflows}
  \small
  \renewcommand{\arraystretch}{1.3}
  \begin{tabular}{>{\centering\arraybackslash}p{1.1cm} p{5.0cm} p{5.0cm} p{5.0cm}}
    \toprule
    \textbf{Phase} & \textbf{A: Density-Based Clustering} & \textbf{B: Partition-Based Clustering} & \textbf{C: Topic Modeling (NMF)} \\
    \midrule
    \textbf{1}\par\smallskip Metrics &
    \emph{Goal:} Find the parameter range where $K$ stabilizes and noise becomes manageable.\newline
    \emph{Watch:} Noise~\% drop then plateau; $K$ stabilize; Silhouette peak; Davies--Bouldin minimum. Complete iterations mark the viable range.\newline
    {\scriptsize[Figs.~\ref{fig:iter1is}; Appendix A, Fig. A.10]} &

    \emph{Goal:} Narrow the candidate $K$ range beyond the SSE elbow.\newline
    \emph{Watch:} Silhouette peak or plateau; Calinski--Harabasz peak; Davies--Bouldin minimum. Metrics may disagree---note where most are favorable and complete iterations cluster. 
    {\scriptsize[Fig.~\ref{fig:pbc_quality}]} &

    \emph{Goal:} Identify where adding topics yields diminishing returns.\newline
    \emph{Watch:} Reconstruction~\% flattening; Coherence ($C_V$) plateau or peak; Global diversity remaining high; Mean exclusivity stable. Complete iterations narrow candidates. 
    {\scriptsize[Fig.~\ref{fig:tm_stability}]} \\
    \midrule
    \textbf{2}\par\smallskip Transitions &
    \emph{Goal:} Identify where all real clusters have formed and persist.\newline
    \emph{Watch:} Noise band shrinking left-to-right; real clusters as persistent horizontal bands. Beyond the stable range, bands converge (over-merging) or fragment (over-splitting).
    {\scriptsize[Figs.~\ref{fig:iter1is},~\ref{fig:iter3_sankey}]} &

    \emph{Goal:} Determine whether splits at $K{+}1$ are meaningful.\newline
    \emph{Watch:} Persistent bands = core structure. A clean binary split = genuine sub-group. Multiple thin bands feeding a new group = incoherent split. Beyond optimal $K$, splits add no structure.
    {\scriptsize[Fig.~\ref{fig:pbc_transitions}]} &

    \emph{Goal:} Distinguish genuine new themes from wastebaskets.\newline
    \emph{Watch:} Persistent bands = stable topics. A new band from a single parent = meaningful differentiation. A new band drawing from many parents = wastebasket collecting generic vocabulary.
    {\scriptsize[Fig.~\ref{fig:tm_sankey}]} \\
    \midrule
    \textbf{3}\par\smallskip Confidence &
    \emph{Goal:} Confirm that non-noise clusters have well-defined boundaries.\newline
    \emph{Watch:} Real clusters: membership near~1, outlier near~0. Noise: membership${\approx}$0, outlier${\approx}$1 (expected). Bimodal violins suggest the cluster conflates sub-populations. 
    {\scriptsize[Appendix~A]} &

    \emph{Goal:} Assess whether increasing $K$ degrades assignment confidence.\newline
    \emph{Watch:} High, compact violins = good separation. Wider or bimodal lower tails at $K{+}n$ signal that additional clusters introduce boundary ambiguity without improving structure. 
    {\scriptsize[Fig.~\ref{fig:pbc_20_24}]} &

    \emph{Goal:} Verify that documents clearly belong to their dominant topic.\newline
    \emph{Watch:} Membership near~1 = clear single-topic documents. High outlier (entropy) = diffuse multi-topic documents. Broadening violins at $K{+}1$ signal over-decomposition. 
    {\scriptsize[Fig.~\ref{fig:tm_sankey}]} \\
    \midrule
    \textbf{4}\par\smallskip Content \& Context &
    \emph{Goal:} Validate clusters against domain knowledge.\newline
    \emph{Actions:} Propagate grouping to map/space-time cube; check geographic and temporal coherence. Use transition-class propagation to isolate borderline members and inspect their domain location.\newline
    {\scriptsize[Figs.~\ref{fig:iter1map},~\ref{fig:iter3_map}]} &

    \emph{Goal:} Confirm distinct, interpretable cluster profiles.\newline
    \emph{Actions:} Parallel coordinates for feature profiles; map for spatial contiguity. Transition-class propagation reveals what members are lost/gained at $K{\pm}1$ and where they go.\newline
    {\scriptsize[Figs.~\ref{fig:pbc_map20},~\ref{fig:pbc_transition_all}]} &

    \emph{Goal:} Verify thematic coherence and external validity.\newline
    \emph{Actions:} Word clouds for gestalt identity; connector tooltips for lost/gained terms; temporal prevalence charts for alignment with known events. Transition word clouds isolate migrating vocabulary.\newline
    {\scriptsize[Figs.~\ref{fig:tm_wc15},~\ref{fig:tm_temporal}]} \\
    \midrule
    \textbf{5}\par\smallskip Archetypes &
    \emph{Goal:} Confirm all persistent patterns are captured.\newline
    \emph{Check:} Noise archetype separate from cluster archetypes; all real archetypes present at selected parameters. Groups should sit close to archetype centroids in 2D. 
    {\scriptsize[Appendix A, Fig. A.11]} &

    \emph{Goal:} Verify robustness to initialization.\newline
    \emph{Check:} All archetypes present $\Rightarrow$ full structure captured. Seed sweep: horizontal Sankey bands and tight 2D clusters confirm negligible drift.\newline
    {\scriptsize[Appendix~B]} &

    \emph{Goal:} Confirm that all thematic archetypes are represented.\newline
    \emph{Check:} All topic archetypes present; high non-noise percentage ($>$85\%) confirms global stability. Noise-assigned topics are idiosyncratic.
    {\scriptsize[Fig.~\ref{fig:tm_stability}]} \\

        \midrule
    \textbf{6}\par\smallskip Decision &
    \emph{Consolidate understanding.} Which spatial patterns are genuine? At what scales? \emph{Select:} $\varepsilon$ and \texttt{min\_samples} where $K$ is stable, noise acceptable, iteration complete, and clusters confirmed by domain views. Sweep the second parameter to verify joint robustness.
    &

    \emph{Consolidate understanding.} What is the demographic typology? Where do boundaries lie? \emph{Select:} The largest $K$ at which all groups are stable, confident, complete, and interpretable. Verify with a seed sweep. Reject higher $K$ if splits lack coherence.
    &

    \emph{Consolidate understanding.} What research communities exist? How do they relate? \emph{Select:} The $K$ where all themes are present, none redundant, all coherent and temporally valid. Reject higher $K$ if new topics are wastebaskets or fragment existing communities.
    \\
    \bottomrule
  \end{tabular}
\end{table*}

\subsection{Common Six-Phase Skeleton}
\label{sec:wf_common}

A preliminary \textbf{Phase~0: Data cleaning} may be needed when initial sweeps reveal stable groups driven by data artifacts---encoding errors, missing-value patterns, duplicate records, or boilerplate text---rather than genuine structure. Their unnatural stability and incoherent profiles make them conspicuous in the IS display.

\textbf{Phase~1: Metric overview and archetype completeness.} Inspect the metrics chart~(T1). Identify candidate ranges where metrics are favorable: elbows, plateaus, peaks, or minima depending on direction. Note which iterations are marked complete (all HDBSCAN archetypes present) and adjust the archetype threshold if needed~(T8).

\textbf{Phase~2: Transition assessment.} Switch to Sankey mode~(T2, T3). Trace group evolution: persistent bands indicate stable groups; diverging bands indicate splits; converging bands indicate merges. Hide irrelevant iterations to focus on the candidate range.

\textbf{Phase~3: Confidence assessment.} Activate violin plots~(T7). Groups with high, concentrated membership and low outlier scores are well-defined; groups with spread-out or bimodal violins may be artifacts or conflating sub-populations.

\textbf{Phase~4: Content and context inspection.} Examine group content~(T4): terms and word clouds for topics; feature profiles via parallel coordinates for clusters. Propagate groupings to domain views~(T5, T6): maps for spatial data, temporal prevalence charts for longitudinal data.

\textbf{Phase~5: Archetype verification.} In the 2D embedding, verify that the selected iteration's groups correspond to distinct, well-separated archetype clusters. Toggle between per-item and per-archetype coloring to check consistency. Optionally sweep random seeds to confirm robustness.

\textbf{Phase~6: Decision.} Consolidate the structural understanding accumulated through Phases~1--5: which groups are genuine, how they relate to each other, and where domain meaning resides. Select the configuration where statistical quality, group stability, membership confidence, archetype completeness, and domain interpretability converge---or articulate what has been learned even when no single configuration is fully satisfactory. If needed, refine the parameter range and iterate.

\subsection{Cross-Workflow Synthesis: Structural Contributions}
\label{sec:wf_synthesis}

Beyond the individual workflows, their synthesis reveals four structural principles that constitute a deeper methodological contribution---they generalize across method families and characterize what it means to \emph{supervise} an unsupervised process.

\textbf{Progressive refinement.} All three scenarios proceeded through multiple computation rounds, each informed by the previous one: coarse-to-fine $\varepsilon$ sweep then \texttt{min\_samples} sweep (Workflow~A); $K$ sweep then seed verification (Workflow~B); data cleaning, full sweep, focused transition inspection (Workflow~C). This pattern is enabled by IS's ability to launch new computation rounds without losing context from previous ones.

\textbf{Converging evidence and cumulative understanding.} No single phase or indicator sufficed for any decision---nor was the decision itself the sole output. In all three scenarios, the exploration process yielded structural insights beyond the final parameter choice: the VAST demonstration revealed the multi-scale density hierarchy of contamination patterns and hospital clusters; the EU demonstration exposed the relationship between demographic archetypes and geographic peripherality; the VIS demonstration uncovered how research communities differentiate as topic granularity increases. The final selection was justified by the \emph{convergence} of evidence across phases, but the knowledge gained \emph{en route}---about transitions, boundary members, and archetype relationships---constitutes an equally important analytical product. This multi-evidence reasoning pattern distinguishes SI from approaches that reduce parameter selection to optimizing one score.

\textbf{Transition-class propagation as a knowledge-generation tool.} Across all workflows, the ability to create temporary class attributes from inter-iteration flows and propagate them to domain views proved essential for understanding \emph{what changes} between parameter settings---and \emph{why}: borderline members occupy cluster edges, revealing the spatial extent of contamination zones (Workflow~A, Fig.~\ref{fig:iter3_map}); the violet cluster's losses are geographically peripheral, exposing the demographic gradient at Europe's northern fringe (Workflow~B, Fig.~\ref{fig:pbc_transition_all}); spatio-temporal vocabulary characterizes migrating documents, clarifying how the VIS community's temporal-analysis subfield relates to spatial analytics (Workflow~C, Fig.~\ref{fig:time_topic_lost_tw}). This mechanism transforms abstract transition flows into domain-grounded knowledge---the analyst learns about the data, not merely about the algorithm's behavior.

\textbf{Rejection as a positive outcome.} The workflows are designed not only to select configurations but to \emph{reject} alternatives with explicit justification. $K = 24$ was rejected because violins revealed boundary ambiguity and the map showed fragmentation without interpretive gain. $n_{\mathrm{topics}} = 16$ was rejected because the Sankey view diagnosed a wastebasket pattern and word clouds confirmed the absence of a coherent identity. Explicit, documented rejection strengthens confidence in the final selection and supports reproducibility.

\subsection{Practical Recommendations}
\label{sec:wf_practical}

The following recommendations generalize from our experience across the three workflows. An extended version with additional guidance is available in the supplementary materials.

\begin{itemize}[nosep,leftmargin=*]
\item \textbf{Start coarse, refine progressively.} A wide sweep at coarse resolution identifies the region of interest; two to three rounds of narrowing typically suffice.
\item \textbf{Let Sankey bands define stability.} Metrics narrow the candidate range, but wide horizontal bands across neighboring configurations provide the definitive persistence evidence.
\item \textbf{Use domain views to break metric ties.} When configurations score similarly, geographic contiguity, temporal coherence, or thematic interpretability discriminate in ways no internal metric can.
\item \textbf{Inspect what changes, not only what persists.} Transition-class propagation is most informative when applied to \emph{losses}---if they are spatially random or semantically incoherent, the finer configuration may over-fit; if they form a coherent sub-population, genuine sub-structure may be emerging.
\item \textbf{Treat archetype completeness as necessary, not sufficient.} Always verify complete iterations through Phases~3--5; conversely, an iteration missing one archetype may be preferable if domain context explains the absence.
\item \textbf{Verify seed robustness} for partition-based methods (20--30 seeds with Sankey inspection).
\item \textbf{Document rejections explicitly.} Recording \emph{why} an alternative was rejected strengthens confidence and reveals data structure.
\end{itemize}

\section{Evaluation}
\label{sec:evaluation}

We evaluate IteraScope through four complementary lenses: (1)~a task-coverage analysis demonstrating comprehensive support for all identified analytical requirements, (2)~qualitative case-study evidence from the three demonstrations, (3)~feedback from professional data scientists, and (4)~feedback from visualization experts. A longitudinal evaluation with MSc data science students is planned for the upcoming semester (Section~\ref{sec:discussion}).

\subsection{Task Coverage and Positioning}
\label{sec:eval_coverage}

Table~\ref{tab:task_coverage} maps each analytical task (T1--T8, from Section~\ref{sec:tasks}) to the IS components that support it and the demonstration sections where it was exercised. All eight tasks are supported by at least two IS components and were demonstrated in at least two of the three use cases. Notably, no single component addresses fewer than two tasks, and no task relies on a single component alone---confirming that the coordinated multi-view design provides redundant, mutually reinforcing support for each analytical need. The table also shows that the three demonstrations collectively exercise the full task set, with each demonstration emphasizing different subsets: the density-based scenario stresses T2 (transitions) and T5 (spatial context); the partition-based scenario stresses T7 (confidence) and T6 (propagation); and the topic-modeling scenario stresses T4 (content inspection) and T8 (archetype verification). This complementary coverage validates that the six-phase workflow skeleton adapts meaningfully to different method families rather than merely relabeling the same steps.

To position these capabilities relative to existing tools, we compare with the most closely related systems. ClusterVision~\cite{kwon2018clustervision} ranks clustering results by quality metrics and supports side-by-side comparison, addressing T1 and partially T4, but does not track how specific groups evolve across configurations (T2, T3) and provides no membership-confidence visualization (T7) or archetype-based completeness assessment (T8). Clustrophile~2~\cite{cavallo2019clustrophile} offers guided iterative refinement with projection-based exploration, covering T1 and T4, and partially supporting workflow structure---but operates on individual results rather than treating the full parameter sequence as an analytical object; it lacks Sankey-style transition tracking, violin plots, and domain-linked propagation (T5, T6). Alexander and Gleicher~\cite{alexander2016task} provide task-driven topic-model comparison with topic alignment across~$K$ values, partially addressing T2 and T4 for topic modeling specifically, but without flow-volume quantification, confidence assessment, or multi-method generality. 
LDAvis~\cite{sievert2014ldavis} supports single-model exploration (T4) through inter-topic projection but does not address parameter selection or cross-configuration comparison at all. ParaClus~\cite{yu2026parallel} and \textsc{Utopian}~\cite{choo2013utopian} share individual design elements (parallel-axis layout and NMF interactivity, respectively) but neither provides sequence-level stability tracking, archetype detection, membership-confidence visualization, or the structured multi-phase workflow that characterizes SI.

The distinguishing capabilities of SI/IS are thus: (1)~sequence-level group tracking with Sankey flows that make splits, merges, and member counts explicit (T2, T3); (2)~HDBSCAN-based recurrent-archetype detection with complete-iteration marking (T8); (3)~globally normalized violin plots for membership confidence (T7); (4)~transition-class propagation that bridges abstract parameter-space patterns with domain-specific views (T5, T6); and (5)~a unified workflow framework supporting three method families. No prior system combines these capabilities in a single coordinated environment.

\begin{table}[tb]
  \centering
  \caption{Task coverage: each analytical task is supported by specific IteraScope components and was demonstrated in multiple use cases. }
  \label{tab:task_coverage}
  \small
  \renewcommand{\arraystretch}{1.15}
  \begin{tabular}{c p{7cm} }
    \toprule
    \textbf{Task} & \textbf{IS Components} \\
    \midrule
    T1 & Metrics chart, tooltips, completeness markers \\
    T2 & Sankey bands, axis hiding, connector tooltips \\
    T3 & 2D embedding, click web, archetype coloring  \\
    T4 & Word clouds, parallel coords, connector tooltips \\
    T5 & Map, space-time cube, temporal prevalence \\
    T6 & Transition-class propagation, domain-linked views \\
    T7 & Violin plots (split/overlay modes) \\
    T8 & Archetype coloring, completeness markers, 2D embedding, HDBSCAN threshold control \\
    \bottomrule
  \end{tabular}
\end{table}

\subsection{Case-Study Evidence}
\label{sec:eval_cases}

The three demonstrations in Sections~\ref{sec:vis_density}--\ref{sec:vis_topics} serve as usage-scenario evaluations~\cite{lam2012empirical,isenberg2013systematic}, exercising the workflows on datasets with different characteristics (spatial, tabular, textual), scales (${\sim}1\,000\,000$ messages, ${\sim}1\,500$ regions, ${\sim}3\,500$ papers), and method families. In each case, the SI workflow led to a well-justified parameter selection that could not have been achieved by inspecting metrics alone:

\begin{itemize}
  \item \emph{Density-based} (Section~\ref{sec:vis_density}): Transition-class propagation revealed that borderline members between \texttt{min\_samples}~$=$~130 and 150 are spatially coherent at cluster edges---information invisible in any scalar metric.
  \item \emph{Partition-based} (Section~\ref{sec:vis_partition}): Violin plots provided the decisive evidence against $K = 24$: boundary ambiguity that the silhouette score alone could not distinguish from $K = 20$.
  \item \emph{Topic modeling} (Section~\ref{sec:vis_topics}): The Sankey view diagnosed the wastebasket pattern at $n_{\mathrm{topics}} = 16$ that coherence metrics failed to flag, while temporal prevalence charts provided external validation unavailable to any internal metric.
\end{itemize}

\noindent In all three cases, the workflow surfaced information that automated metric-based selection would miss, supporting our central claim that human-in-the-loop supervision adds irreplaceable analytical value to iterative clustering and topic modeling.

\subsection{Expert Feedback: Data Scientists}
\label{sec:eval_ds}

We conducted informal evaluation sessions with $N=3$  data scientists ($\ge10$ years experience) who regularly apply clustering and topic modeling in their work. Participants explored the IS display on two of the three demonstration datasets (partition-based clustering and topic modeling) and were asked to think aloud while navigating the six-phase workflow.

\textbf{Positive reception.} Participants expressed strong appreciation for the overall approach. They valued the ability to make \emph{informed decisions} grounded in multi-faceted evidence rather than relying on a single score or elbow heuristic. The consistent color assignment across iterations was specifically praised: participants noted that tracking a group visually across a $K$-sweep without needing to match labels manually removed a significant cognitive burden. The concept of recurrent archetypes was considered ``excellent''---participants found it intuitive that genuine structure should reappear across many parameter settings, and the completeness markers provided an immediate, actionable signal.

\textbf{Identified issue: HDBSCAN threshold sensitivity.} Participants identified one significant issue in the evaluated implementation: the selection of the \texttt{min\_cluster\_size} parameter for HDBSCAN archetype detection was not informed by any visual guidance. Changing this threshold alters the number of detected archetypes, which in turn changes the set of archetype centroids included in the MDS embedding used for color assignment. Consequently, adjusting the threshold could cause colors to shift unpredictably---undermining the very consistency that participants valued.

\textbf{Resulting improvement.} This feedback motivated a redesign of the archetype-detection interface. The system now runs HDBSCAN iteratively for all plausible \texttt{min\_cluster\_size} values (from 2 to $N_{\text{iterations}} - 1$) and visualizes the number of detected archetypes and the percentage of noise-assigned groups as functions of the threshold. This chart enables the analyst to make an informed choice---for example, selecting a threshold at a stability plateau where small changes do not alter the archetype count. Furthermore, the results of \emph{all} HDBSCAN runs are included in the MDS embedding together with all group-level feature vectors from all iterations. Because the embedding is computed once over this comprehensive set, color assignment remains consistent regardless of which \texttt{min\_cluster\_size} the analyst subsequently selects---resolving the instability that participants had identified.

\subsection{Expert Feedback: Visualization Researchers}
\label{sec:eval_vis}

A separate evaluation session was conducted with a group of visualization experts. Participants were shown the full IS display across all three demonstration datasets and invited to critique both the visual design and the workflow logic.

\textbf{Positive reception.} The experts considered the idea of treating iterative grouping results as a first-class analytical object to be both novel and useful. They noted that the implementation supports building multiple pairs of 1D/2D embeddings (produced using different projection methods with different parameters) at once and switching between them interactively---a feature that enables the analyst to assess whether observed patterns are projection artifacts or genuine structural features.

\textbf{Recommendation: Rank-by-feature for embedding selection.} The experts recommended applying a rank-by-feature framework~\cite{seo2005rank} to the embedding-selection problem: rather than requiring the analyst to manually compare all available projections, the system could rank 1D/2D embedding variants by quality criteria (e.g., trustworthiness, continuity, neighborhood preservation) and suggest the most expressive variants. This would reduce the combinatorial burden of choosing among projection methods and parameter settings---particularly relevant when the analyst is unfamiliar with the relative strengths of t-SNE, UMAP, PaCMAP, and LocalMAP for a given dataset. We consider this an important direction for future work that would further reduce analyst effort in Phases~2 and~5 of the workflow.

\section{Discussion and Limitations}
\label{sec:discussion}

The demonstrations and expert evaluations confirm that the SI approach delivers actionable guidance for parameter selection across three method families. At the same time, they exposed design trade-offs and open challenges that merit discussion.

\textbf{Generalizability.} The workflow skeleton (Section~\ref{sec:wf_common}) applies to any unsupervised method that assigns items to groups; method-specific indicators can be substituted while the phases remain the same. The workflows have been demonstrated on three method families. 
Extension to hierarchical clustering, fuzzy c-means, or neural topic models (e.g.\ BERTopic~\cite{grootendorst2022bertopic}) would require adapting the membership/outlier semantics and term-level tooltip design, but the six-phase skeleton and transition-assessment logic should transfer directly. In particular, NMF produces non-negative, sparse factors that yield naturally interpretable topics and a straightforward membership probability ($\max(w_i)/\sum w_i$); for LDA~\cite{blei2003latent}, the Dirichlet-distributed document--topic mixtures may produce less peaked membership distributions, while BERTopic's embedding-based assignments lack an explicit per-document topic mixture altogether---requiring alternative confidence indicators (e.g., cosine similarity to the nearest topic centroid) that the violin-plot framework can accommodate without structural changes.

\textbf{Scalability.} Parallel execution scales computation near-linearly with cores. The visual bottleneck is Sankey readability: beyond ${\sim}50$ iterations or ${\sim}50$ groups per iteration, bands become cluttered. Axis hiding, 2D embedding focus, and filtering by transition strength partially mitigate this, but hierarchical aggregation of similar iterations remains future work.

\textbf{HDBSCAN threshold sensitivity.} Archetype detection depends on the \texttt{min\_cluster\_size} parameter; different thresholds yield different archetype counts. The evaluation with data scientists (Section~\ref{sec:eval_ds}) identified this as a concern and motivated the iterative HDBSCAN sweep interface, which visualizes archetype count and noise percentage as functions of the threshold and ensures consistent color assignment by including all runs in the shared MDS embedding. The analyst can now make an informed threshold choice at a stability plateau and observe how the set of complete iterations changes.

\textbf{Stability $\neq$ correctness.} A stable group is not necessarily a ``true'' group: artifacts of feature engineering or preprocessing can also persist across parameter configurations. Domain views (Phase~4) serve as a crucial validation mechanism---as demonstrated by the geographic confirmation in Section~\ref{sec:vis_density} and the temporal validation in Section~\ref{sec:vis_topics}.

\textbf{Iteration reordering and grouping.} When the sweep parameter is a random seed rather than a meaningful ordinal quantity (e.g., $K$ or $\varepsilon$), the left-to-right axis order in the Sankey view is arbitrary and the resulting transition bands may be difficult to interpret. A natural extension is to let the analyst reorder iterations according to a selected quality indicator (e.g., silhouette, coherence), a pairwise similarity measure, or a 1D embedding of the iteration-level distance matrix. Beyond reordering, clustering or embedding the iterations themselves---treating each full grouping result as a single high-dimensional object---could reveal families of structurally similar solutions, helping the analyst identify consensus groups of runs and outlier initializations without inspecting every seed individually.

\textbf{Workflow rigidity.} The six-phase structure is a guideline, not a rigid protocol. Experienced analysts may skip or reorder phases; the system supports free exploration at any time. The demonstrations themselves illustrate non-sequential use: Section~\ref{sec:vis_density} visited Phase~4 before Phase~2, and Section~\ref{sec:vis_partition} interleaved transition inspection with confidence assessment.

\textbf{Learning curve.} The system exposes many interaction modalities (hover, click, right-click context menus, axis toggling, violin mode switching). While the progressive demonstration structure introduces complexity gradually, a formal learnability study with novice users remains outstanding. The planned longitudinal classroom evaluation (below) will provide initial evidence on learning barriers.

\textbf{Automation potential.} Stability scores (e.g.\ Jaccard persistence across consecutive iterations) could be computed automatically and used to pre-highlight candidate configurations. The HDBSCAN archetype completeness check is already a step in this direction. The rank-by-feature framework for embedding selection recommended by visualization experts (Section~\ref{sec:eval_vis}) represents another avenue for automated guidance. We see these as complementary: automated suggestions seed the visual exploration rather than replacing it.

\textbf{Formal user study.} The present evaluation relies on case-study evidence, task-coverage analysis, and informal expert feedback. A controlled comparative study---measuring task completion time and insight quality with vs.\ without IS support---would strengthen the empirical grounding and is a priority for future work, ideally conducted with domain analysts from fields where clustering decisions carry operational consequences (e.g.\ epidemiology, urban planning, digital humanities).

\textbf{Longitudinal classroom evaluation.} Beginning in the upcoming semester, MSc students in a data science program will use the Google Colab notebooks as part of their coursework and diploma research. This deployment will provide evidence on learnability, workflow adoption, and the extent to which structured parameter supervision improves analytical conclusions compared to traditional single-metric approaches.

\section{Conclusion}
\label{sec:conclusion}

We presented SmartIterator, a visual analytics approach for supervising the parameter-selection process in unsupervised data grouping, operationalized through the IteraScope coordinated display. Our main contribution is three structured workflows---for density-based clustering, partition-based clustering, and NMF topic modeling---each guiding the analyst through six phases that combine quality metrics, transition stability, membership confidence, content inspection, recurrent-archetype verification, and domain contextualization. A cross-workflow synthesis distills four generalizable principles: progressive refinement, converging evidence, transition-class propagation as a knowledge-generation tool, and explicit rejection as a positive analytical outcome.

Demonstrations on the VAST Challenge 2011 (with ground-truth validation), EU NUTS-3 population data, and 30~years of IEEE VIS papers show that the workflows lead to well-justified, interpretable parameter choices across diverse data types and scales. Expert feedback from data scientists and visualization researchers confirmed the approach's practical value and drove targeted improvements to the archetype-detection interface.

The key takeaway is that treating the \emph{sequence of grouping results} as a first-class analytical object---and providing structured, method-specific guidance for its exploration---generates knowledge about data structure that no single ``best'' result can provide. The analyst does not merely pick a number; the analyst \emph{learns} about the data by studying how structure emerges, persists, and transforms across configurations, and arrives at a decision grounded in that accumulated understanding.



\bibliographystyle{abbrv-doi-hyperref}
\bibliography{references}

\end{document}